\documentclass{aa}  

\usepackage{graphicx}
\usepackage{txfonts}
\usepackage[version=3]{mhchem}
\usepackage{xcolor}
\usepackage{lscape}
\usepackage{color, colortbl}
\usepackage{arydshln}

\usepackage{booktabs}
\usepackage{hyperref}
\usepackage{amssymb,amsmath,enumitem,longtable,color}
\usepackage{multirow}

\newcommand{\micron}{\textmu m\xspace}

\usepackage{ulem}

\begin{document}

   \title{Metallic species in interstellar medium: Astrochemical modeling}


   \author{Ankan Das\inst{\ref{inst1},\ref{inst2}} \and Milan Sil\inst{\ref{inst3},\ref{inst4}} \and Paola Caselli\inst{\ref{inst1}}}

   \institute{Max-Planck-Institute for extraterrestrial Physics, P.O. Box 1312 85741 Garching, Germany
             \label{inst1}
\and
   Institute of Astronomy Space and Earth Science, P177 CIT Road, Scheme 7m, Kolkata 700054, India \\
   \email{ankan.das@gmail.com}
              \label{inst2}
        \and
              Univ. Grenoble Alpes, CNRS, IPAG, 38000 Grenoble, France \\
              \email{milansil93@gmail.com}
              \label{inst3}
        \and
             Univ Rennes, CNRS, IPR (Institut de Physique de Rennes) - UMR 6251, F-35000 Rennes, France
             \label{inst4}}


 
  \abstract {
Metal-bearing species in diffuse or molecular clouds are often overlooked in astrochemical modeling except for the charge exchange process. However, catalytic cycles involving these metals can affect the abundance of other compounds. We prepared a comprehensive chemical network for Na, Mg, Al, Fe, K, and Si-containing species. Assuming water as the major constituent of interstellar ice in dark clouds, quantum chemical calculations were carried out to estimate the binding energy of important metallic species, considering amorphous solid water as substrate. Significantly lower binding energies (approximately 5 to 6 times) were observed for Na and Mg, while the value for Fe was roughly 4 times higher than what was used previously.  Here, we calculated binding energy values for Al and K, for which no prior guesses were available. The obtained binding energies are directly implemented into the models of diverse interstellar environments. The total dipole moments and enthalpies of formation for several newly included species are unknown. Furthermore, the assessment of reaction enthalpies is necessary to evaluate the feasibility of the new reactions under interstellar conditions. These parameters were estimated and subsequently integrated into models.
Some additional species that were not included in the UMIST/KIDA database have been introduced. The addition of these new species, along with their corresponding reactions, appears to significantly affect the abundances of related species.
Some key reactions that significantly influence general metal-related chemistry include: \( \text{M}^+ + \text{H}_2 \rightarrow \text{MH}_2^+ + h\nu \), \( \text{MH} + \text{O} \rightarrow \text{MO} + \text{H} \) (M = Fe, Na, Mg, Al, or K),  and $\rm{M_1^+ + M_2H \rightarrow M_1H + M_2^+}$ (where $\rm{M_1 \neq M_2}$, M$_1$,  M$_2=$ Na, Mg, Al, K, Fe). These reactions have a notable impact on the abundance of these species. Significant changes were observed in magnesium and sodium-bearing cyanides, isocyanides, and aluminum fluoride when additional reaction pathways were considered.
}


   \maketitle

\section{Introduction} \label{sec:intro}
Almost 90\% of all refractory elements condense in the innermost envelopes of evolved stars. The condensation occurs at the condensation temperature of the ingredients. So, it is expected that dust grains are layered based on the highest-lowest condensation sequence in the following order: Al, Ti, Ca, Fe, (Mg, Si), P, (Na, K), and S \citep{turn92p}.
The chemistry associated with the metals is important in magnetized clouds because metals play a crucial role in regulating the ionization fraction of the interstellar medium (ISM).
However, their identification in space is limited due to the lack of spectroscopic information and weak transitions in the radio domain. Nevertheless, a handful of metallic compounds have been identified to date. It includes multiple metallic halides \citep[NaCl, AlCl, KCl, and AlF;][]{cern87},
metal cyanides \citep[MgCN, KCN, SiCN, NaCN, FeCN, MgC$_3$N;][]{ziur95,pull10,guel00,turn94,zack11,cern19b}, metal isocyanides \citep[MgNC, CaNC, SiNC, AlNC, HMgNC;][]{guel93,cern19a,guel04,ziur02,cabe13}, etc.
The IRC+10216, circumstellar envelope (CSE) of the carbon-rich star, is a major inventory of these species found to date. In addition, some species were also located toward the more evolved carbon star CRL 2688 \citep{high03}. 

Metal oxides can be detected in the atmosphere of stars. However, it is difficult to identify refractory elements, like Mg, Fe, Ca, and Si, in molecular clouds due to their high depletion. While SiO has been observed along multiple lines of sight, its abundance is several orders of magnitude lower than the solar abundance of Si. Typically, SiO can be found in regions of shocks or outflows. The absorption of the $J=5-4$ transition of FeO has been observed towards Sgr B2(M) by \citet{walm02} and \citet{furu03}.

Recently, NaCl has been detected in the massive protostellar system, Orion Source I disk \citep{Gins19,wrig20}.
Also, \citet{tana20} found NaCl, SiO, and SiS emissions in the inner disk ($\sim 100$ au scale) of a massive protostellar disk, IRAS 16547-4247. Most interestingly, all these emissions do not necessarily have very high upper-state energy, and critical densities are also not too high ($10^5-10^6$ cm$^{-3}$). Identifying these refractory species in the innermost region of massive protostellar disks may be the product of dust destruction, which may be feasible around the inner disk.
Moreover, Al, Na, and K metal-bearing species are closed-shell molecules, mainly obtained in the hotter part of the envelope. On the contrary, Mg, Si, and Fe-related species are mostly open-shell radicals that may react in cold environments. 

The C-rich envelope of evolved stars is well explored by earlier studies \citep{cern12,agun13}.
Here, we focus mainly on metallic compounds in the ISM. 
We studied various star-forming regions where the formation of metallic species is often overlooked. Our focus includes harsh environments that are conducive to molecular formation, such as the H\,{\sc ii}-photodissociation region (PDR). We also examined radiation-shielded areas that are favorable for forming complex species. These areas include diffuse clouds, dark clouds, and hot core regions. A large chemical network for the metallic species is prepared and then coupled with the existing chemical network. Quantum chemical calculations were conducted to study various chemical properties (reaction enthalpies, binding energies, dipole moments etc.). The examination of metal-bearing species in other star-forming regions falls outside the scope of this paper; however, the pathways and chemical parameters presented here are likely applicable in a broader context.

{This paper is organized as follows. First, in Sect.~\ref{sec:physchem}, we explain the results of physical chemistry, including the estimation of binding energies, the molecular dipole moments of the species studied, and the reaction enthalpies for the newly included reactions. 
In Sect.~\ref{sec:chem-route}, we discuss the chemical routes for gas and ice-phase metallic compounds. Sect.~\ref{sec:models} presents an in-depth discussion of the various chemical models used in this work. 
Sect.~\ref{sec:astro} covers the results and their astrophysical implications. Finally, we conclude in Sect.~\ref{sec:con}.}


\begin{table*}
\caption{Computed binding energies of metal-bearing species considering water monomer (default) as the substrate while incorporating a scaling factor of $1.416$ (see text for details). 
\label{tab:be}}
\resizebox{\linewidth}{!}{\begin{tabular}{|cc|cc|}
\hline
			\textbf{Species (Ground state)}& \textbf{Binding energy [K]}	& \textbf{Species (Ground state)} & \textbf{Binding energy [K]}	 \\
			\hline
   \multicolumn{2}{c}{\bf Si-related} & \multicolumn{2}{c}{\bf Na-related} \\
   \hline
  Si (triplet)  & 7187 (6956$^1$, $11\,600\pm3480^2$, $2400\pm500^4$) &  Na (doublet)  & $2408^*$ (2214$^1$, 11\,800$^3$, $10\,600\pm500^4$) \\
       SiH (doublet) & 9560 (8988$^1$, $13\,000\pm3900^2$, $2620\pm500^4$) &  NaH (singlet)  & 7428 (12\,250$^3$) \\
       SiO (singlet)  & 4136$^*$ (3500$^3$, $3150\pm500^4$) & NaO  (doublet) & 8263 \\
       SiC (triplet) & $6022^*$ (5850$^1$, 3500$^3$, $3150\pm500^4$) & NaC (quartet) & 11707 \\
       SiN  (doublet) & 8526 (3500$^3$, $3500\pm500^4$) & NaN  (triplet) & 15438 \\
       SiH$_2$ (singlet) & 11\,329 (3600$^3$, $3190\pm500^4$) & NaS  (doublet) & 12701 \\
       SiH$_3$ (doublet) &  $538^*/726^*$ (1269$^1$, 4050$^3$, $3440\pm500^4$) &  NaCl  (singlet) & { 15019} \\
       SiH$_4$ (singlet) & $842^*$ (1527$^1$, 4500$^3$, $3690\pm500^4$) & NaOH  (singlet) & 16\,609 (14\,650$^3$) \\
       SiOH (doublet) & $5236^*$ & NaCN (singlet) & 14\,770 \\
       SiCN (doublet) & 13\,188 &  NaNC (singlet)  & 9611 \\
       SiNC (doublet) & 10\,969 (4300$^3$, $1350\pm500^4$) & NaC$_2$H (singlet) & --- \\
       SiS (singlet) & $2145^*$ (3800$^3$, $3400\pm500^4$) & NaC$_4$H (singlet) & 1833 \\
       HCSi (doublet) & 530 (3625$^3$, $1050\pm500^4$) & & \\
       HNSi (singlet) & 3110 (5078$^3$, $1100\pm500^4$) & & \\
       \hline
       \multicolumn{2}{c}{\bf Mg-related} & \multicolumn{2}{c}{\bf Fe-related} \\
   \hline
       Mg (singlet) & $860^*$ (776$^1$, 5300$^3$) &  Fe (quintet)  & 16677 (4200$^3$, $3750\pm500^4$) \\
       MgH (doublet) & 8349 (5750$^3$) & FeH (quartet)  & 11\,832 (4650$^3$) \\
       MgH$_2$ (singlet)  & 11\,267 (6200$^3$) & & \\
       MgO (triplet) & 11\,107 & FeO (quintet) & 36\,420 \\
       MgC (triplet) & 1922 & FeC (triplet) & 7684 \\
       MgN (quartet) & 8961 & FeN  (quartet) & --- \\
       MgS  (singlet) & 16\,222 & FeS  (quintet) & 7011 \\
       MgOH (doublet) & 12\,753 & FeOH  (quartet) & --- \\
       MgCN  (doublet) & 13\,196 & & \\
       MgNC (doublet) &  chemisorption & & \\
       HMgNC (doublet) & 16\,348 & & \\
       MgC$_2$H (doublet) & 1834 & & \\
    MgC$_4$H (doublet) & 2566 & & \\

       \hline
       \multicolumn{2}{c}{\bf K-related} & \multicolumn{2}{c}{\bf Al-related} \\
   \hline
              K (doublet) & $2133^*$ & Al (doublet) & 4381 \\
              KH (singlet) & chemisorption & AlH (singlet) & 6624 \\
               & & AlO (doublet) &16233 \\
              KCl (singlet) & chemisorption &  AlCl (singlet)  & 6351 \\
              KCN (singlet) & 6813 & AlCN (singlet) & 9375 \\
              && AlNC (singlet) & 8729 \\
 			\hline
		\end{tabular}}
	\tablefoot{\footnotesize
 $^*$ ZPE-corrected binding energy considering water tetramer as the substrate using MP2/def2-TZVPP level of theory, \\
 {$^1$} ZPE-uncorrected binding energy obtained by \citet{das18} for water tetramer as the substrate using MP2/aug-cc-pVDZ level of theory, \\
{$^2$} \citet{wake17}, \\
{$^3$} Noted in KIDA from the OSU database group of Eric Herbst in 2006, \\
{$^4$} \citet{pent17}.}

\end{table*}
{
\section{Physical chemistry results \label{sec:physchem}}}
\subsection{Binding energy\label{sec:BE}}
 Among the six metallic elements (Na, Mg, Si, Al, K, and Fe) considered here,
the binding energy calculation of only Si was available \citep[$\sim 11,600\pm3480$~K;][]{wake17}. For Na, Mg, and Fe, the KIDA database noted the binding energy as 11\,800~K, 5300~K, and 4200~K, respectively, estimated from very early studies (the OSU database of Eric Herbst in 2006).
However, no follow-up studies supported these estimates via quantum chemical calculations or experiments \citep{pent17,das18}.
\cite{das18} obtained binding energies on water tetramer using MP2/aug-cc-pVDZ level of theory without considering ZPE-correction.
\cite{pent17} provided average binding energies based on the work of \cite{hase93} and \cite{aika96} with an uncertainty set to 500~K.

Since the depletion of heavy elements could control the ionization of the medium and, eventually, the star formation rate, it is beneficial to have a better understanding of the binding energy of these metals.
Here, we perform the quantum chemical calculations to determine the binding energy of some of these key species noted in Table~\ref{tab:be}. We use the MP2/aug-cc-pVDZ level of theory \citep{dunn89} without considering ZPE correction \citep{das18,das21,sil17} for BE computations with water monomer. A scaling factor of $1.416$ \citep[in case of monomer;][]{das18} is used.  Since the aug-cc-pVDZ basis set is not supported for the K-related species, we use the MP2/Def2-TZVPP level of theory \citep{weig05} without considering ZPE corrections (same scaling factor of $1.416$ is used). For water tetramer surfaces (in a few cases), we use the MP2/Def2-TZVPP level of theory considering ZPE corrections to predict the binding energies (no scaling factor is used).
We must admit that the vast astrophysical environments modeled here would not have water as a substrate, and the interaction with other substrates would differ significantly. However, a systematic study based on a single substrate would be better than arbitrary guesses. The computation of binding energy with other substrates is out of the scope of this paper and will be reported elsewhere.
First, the ground state spin multiplicity of the species is verified by reviewing the optimized energies. Then, the ground state spin multiplicity considered in this study is noted in Table~\ref{tab:be}.
Our calculated binding energy of Si of 7187~K with water  monomer is toward the lower limit range provided by \citet{wake17}.
Interestingly, we obtained very low binding energy values for Na (2408~K) and Mg (860~K) with the water tetramer compared to the available literature values.
For the other metals, Al, Fe, and K, we obtained binding energy values of 4381~K, 16\,677~K, and 2133~K, respectively.
We propose calculated binding energies for the elements Al and K, for which no estimation was previously available.
We use a diffusion ($E_b$) to desorption ($E_D$) energy ratio of 0.5 for all the models considered here.

\subsection{Reaction enthalpies \label{sec:enthalpy}}
Apart from the standard gas-phase reactions \citep{mill22}, Table~A.1 (available on Zenodo) depicts some additional gas-phase reactions studied in this work. The reaction enthalpies [$\Delta_r H^0 (298 \ K)$] for some reactions are calculated to evaluate their validity under interstellar conditions.
Although these reaction enthalpies are computed at 298~K, we note that the differences in reaction enthalpies between 298~K and typical dark cloud temperatures ($\sim10$~K) are expected to be negligible, as enthalpy is only weakly temperature-dependent over this range. The main factors influencing reactivity in such low-temperature environments are instead kinetic (e.g., activation barriers) rather than thermodynamic.
The exothermic reactions are highlighted in Table~A.1 (available on Zenodo) and are finally included in our gas-phase chemical network.
The typical method for calculating reaction enthalpies involves determining the heat of formation and applying the appropriate sums and differences mentioned below:
\begin{equation}
    \Delta_rH^0 (298 \ K) = \sum_{products}\Delta_fH^0_{prod}(298 \ K) - \sum_{reactants}\Delta_fH^0_{react}(298 \ K).
\end{equation}
The atomization energy of molecules is calculated using the DFT-B3LYP/6-311+G(d,p) method to determine the enthalpy of formation. All the quantum chemical calculations conducted in this work are performed utilizing the Gaussian 16 computational program \citep{fris16}. Experimental data for the heat of formation of atoms is obtained from \citet{curt97}. Table~\ref{tab:enthalpy} summarizes the standard enthalpy of formation [$\Delta_r H^0 (298 \ K)$] for the species considered. Most values are sourced from the NIST Chemistry WebBook, SRD 69 gas-phase thermochemistry database.
However, in cases where the enthalpies of formation are not available in the literature, or we could not determine it due to lack of experimental data of constituent atoms, we simply took the difference of the sums of electronic and thermal enthalpies for the reactants and the products calculated using the DFT-B3LYP/6-311+G(d,p) level of theory.

{ 
\subsection {Dipole moment}
Ion-neutral reactions play a crucial role in the ISM \citep{herb73}. These reactions often proceed rapidly at low temperatures, influencing the chemical evolution of molecular clouds.

We used the approximated \cite{su82} formula as prescribed by \cite{mill22} to calculate ion-neutral (IN) rate coefficients: 

\begin{equation}
    k= 3.87 \times 10^{-9} \mu_D \mu^{-1/2} \left(\frac{T}{300}\right)^{-1/2} \ {\rm cm^{3} s^{-1},}
\end{equation}
where $\mu_D$ is the electric dipole moment of the neutral species in Debye, and $\mu$ is the reduced mass of the reactants in amu. Dipole moments for some of the metallic species considered in our network were not available, and thus, we employed quantum chemical calculations to estimate it with the DFT-B3LYP/6-311+G(d,p) level of theory. Table \ref{tab:dipole} denotes the dipole moment used in our calculations.}

\begin{table*}
\scriptsize{ 
\caption{ Total dipole moment for the species considered in our model. 
\label{tab:dipole}}
\resizebox{\linewidth}{!}{\begin{tabular}{|cc|cc|cc|}
\hline
			\textbf{Species}& \textbf{Total dipole moment [Debye]}	& \textbf{Species} & \textbf{Total dipole moment [Debye]}& \textbf{Species}& \textbf{Total dipole moment [Debye]}	 \\
			\hline
   \multicolumn{2}{c}{\bf Si-related} & \multicolumn{2}{c}{\bf Na-related} & \multicolumn{2}{c}{\bf Mg-related} \\
   \hline
       SiH  & 0.09$^1$ &  NaH   &  6.02$^2$& MgH  & 1.34$^2$ \\
       SiO   & 3.09 $^1$& NaO   &  8.05$^2$&MgO  &  6.04$^1$\\
       SiC &1.7$^1$ & NaC  &  7.48$^2$& MgC  &  0.67$^2$\\
       SiN  &  2.56$^1$ & NaN   &  6.35$^2$&MgN  &  2.5 $^2$\\
       SiH$_2$  & 0.18$^1$ & NaS  &  8.45$^2$&MgH$_2$   &  0$^2$\\
       SiH$_3$ & 0.01$^2$&  NaCl   &  9.00 $^1$&MgOH  &  1.4 $^1$\\
       SiH$_4$  &0$^1$  & NaOH   &  6.66 $^2$&MgCN   &  5.3$^1$\\
       SiOH  & 1.54$^2$ & NaCN  &  8.85$^1$&MgNC  &  5.22$^1$\\
       SiCN  &3.14$^2$ &  NaNC   & 10.9$^2$&MgS   &  7.66$^2$\\
       SiNC  & 2.00$^1$ &  NaC$_2$H & 7.65 $^2$&MgC$_2$  &  7.9$^1$ \\
       SiS  & 1.74$^1$ & NaC$_4$H  &  9.18$^2$&MgC$_2$H  & 1.68 $^1$\\
       HCSi  & 0.07$^2$ & & &HMgNC  & 3.49$^1$\\
       SiCH$_2$  &0.18$^1$  & &&MgC$_4$H  &  2.12$^1$ \\
       SiCH$_3$  & 0.65$^1$ & & &MgC$_3$N  &  6.38$^1$\\
       HNSi  & 0.25$^1$ & & &MgC$_6$H& 2.5$^1$\\
       HSiO  & 3.39$^1$ & & &MgC$_5$N&7.3$^1$\\
       H$_2$SiO  & 3.82$^1$ & & &MgC$_8$H  &  2.9$^1$\\
       SiC$_2$  & 2.39$^1$ & & &MgC$_7$N  & 8.3 $^1$\\
       SiC$_2$H  & 1.4$^1$ & & &&\\
       SiC$_2$H$_2$  & 0.92$^1$  & & & &\\
       SiO$_2$  & 0.5$^1$ & & &&\\
       HSiS  &  2.06$^1$& & &&\\
       H$_2$SiS  & 2.67$^1$ & & &&\\
       SiC$_3$  & 4.9$^1$ & & &&\\
       SiC$_3$H  & 0.77$^1$ & & &&\\
       SiC$_4$  & 6.42$^1$ & & &&\\
       SiCN  & 3.14$^1$ & & &&\\
       \hline
       \multicolumn{2}{c}{\bf Fe-related} & \multicolumn{2}{c}{\bf K-related}& \multicolumn{2}{c}{\bf Al-related} \\
   \hline
    FeH   & 2.73 $^2$ &KH  & 7.55 $^2$ & AlH  & 0.19 $^1$ \\
     FeO  &  4.08 $^1$&KCl  & 10.58 $^2$     & AlO  & 4.39$^1$\\
     FeC  &  2.71 $^2$ &KCN&12.39$^2$&AlOH &1.04 $^1$\\
     FeN   &  3.9 $^2$&&&AlCl   & 1.63 $^1$\\
      FeS   &  5.68 $^2$&&&AlCN  &  3.74 $^1$\\
      FeOH   & 2.71 $^2$ &         &&AlF&1.44 $^1$\\
      &&&&AlNC&3.31 $^1$  \\
       \hline
		\end{tabular}}}
	\tablefoot{\footnotesize $^1$ \cite{mill22}, $^2$ This work considering DFT-B3LYP/6-311+G(d,p) level of theory.}
\end{table*}

\section{Chemical routes \label{sec:chem-route}}

This study uses two state-of-the-art chemical models: The \textsc{Cloudy} photoionization code version 22.01 \citep{ferl17,shaw22}, and two phase Chemical Model for Molecular Cloud code \citep[CMMC;][]{das15a,sil18,das19,das21,sil21,bhat22,gora17a,gora20,mond21,ghos22,sriv22,rama24}.

\subsection{Gas-phase pathways}
Some essential charge transfer reactions related to Na, Mg, and Si were already considered in the \textsc{Cloudy} code from \cite{mcel13}. However, the CMMC code considered all the charge transfer reactions of these species noted in the UMIST-2022 database for Astrochemistry \citep{mill22}.
Fe-related charge transfer reactions were not considered in the \textsc{Cloudy} code, but they are considered in the CMMC code from UMIST.
Presently, we use the charge transfer reactions of these species used in \cite{mill22} in both codes.
Our study includes additional Al and K-related charge exchange reactions by considering a similar charge exchange reaction analogous to Fe and Na, respectively.
Since not too many negative ions were considered in the \textsc{Cloudy} network, we only limited our consideration to some essential negative ions. However, we have already considered all the reactions with the negative ions from UMIST-2022 in the CMMC code for Na, Mg, Fe, and Si.
Table~A.1 (available on Zenodo) presents the newly included gas-phase reactions in this study. 
As discussed in Sect.~{\ref{sec:enthalpy}}, we calculate the enthalpies of several reactions; however, only exothermic reactions are included in our gas phase network. Furthermore, the adopted rate and the proxy metal used instead are clearly highlighted for reference.

\begin{figure}
    \centering
    \includegraphics[width=0.30\textwidth,angle=-90]{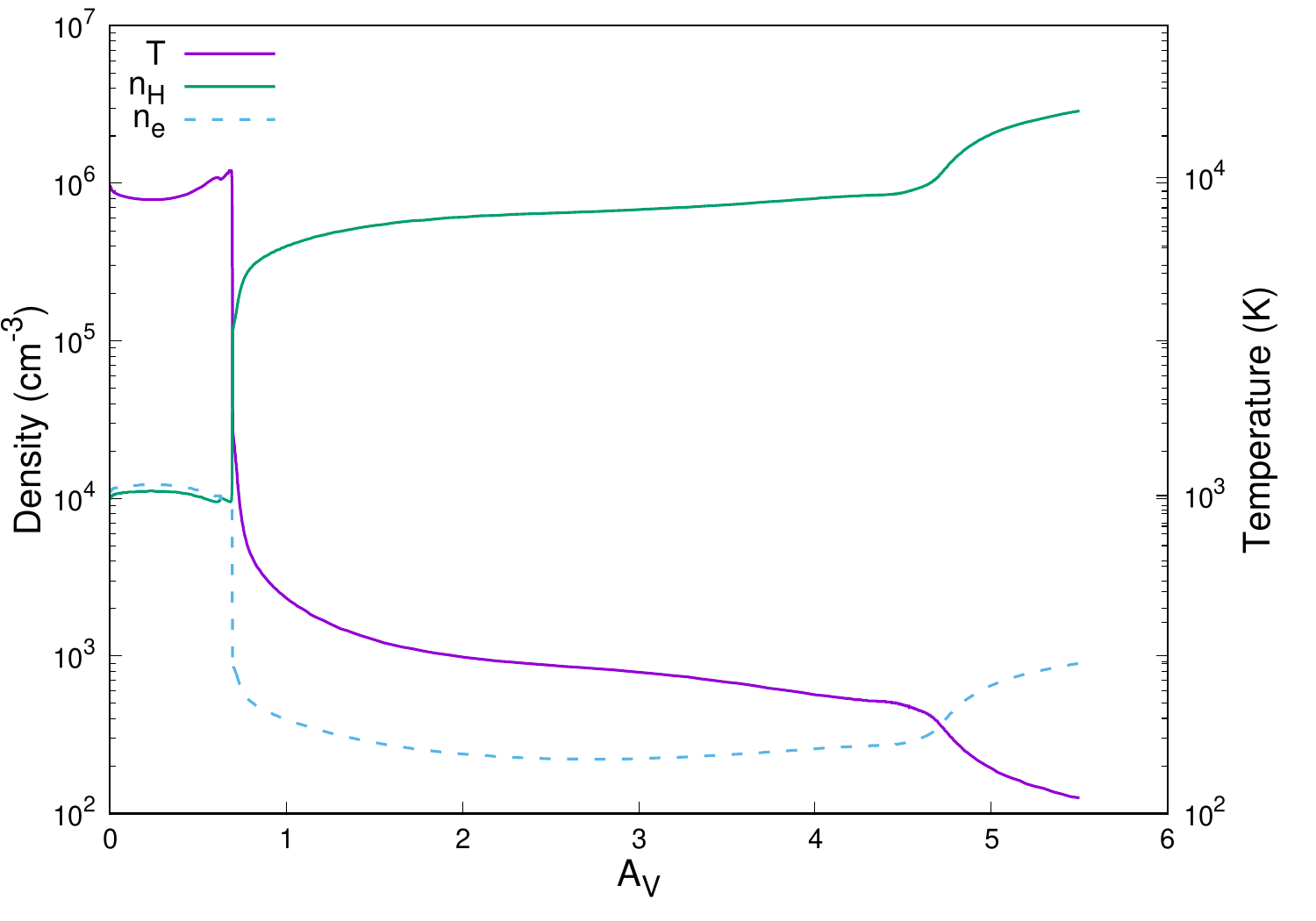}
   \includegraphics[width=0.30\textwidth,angle=-90]{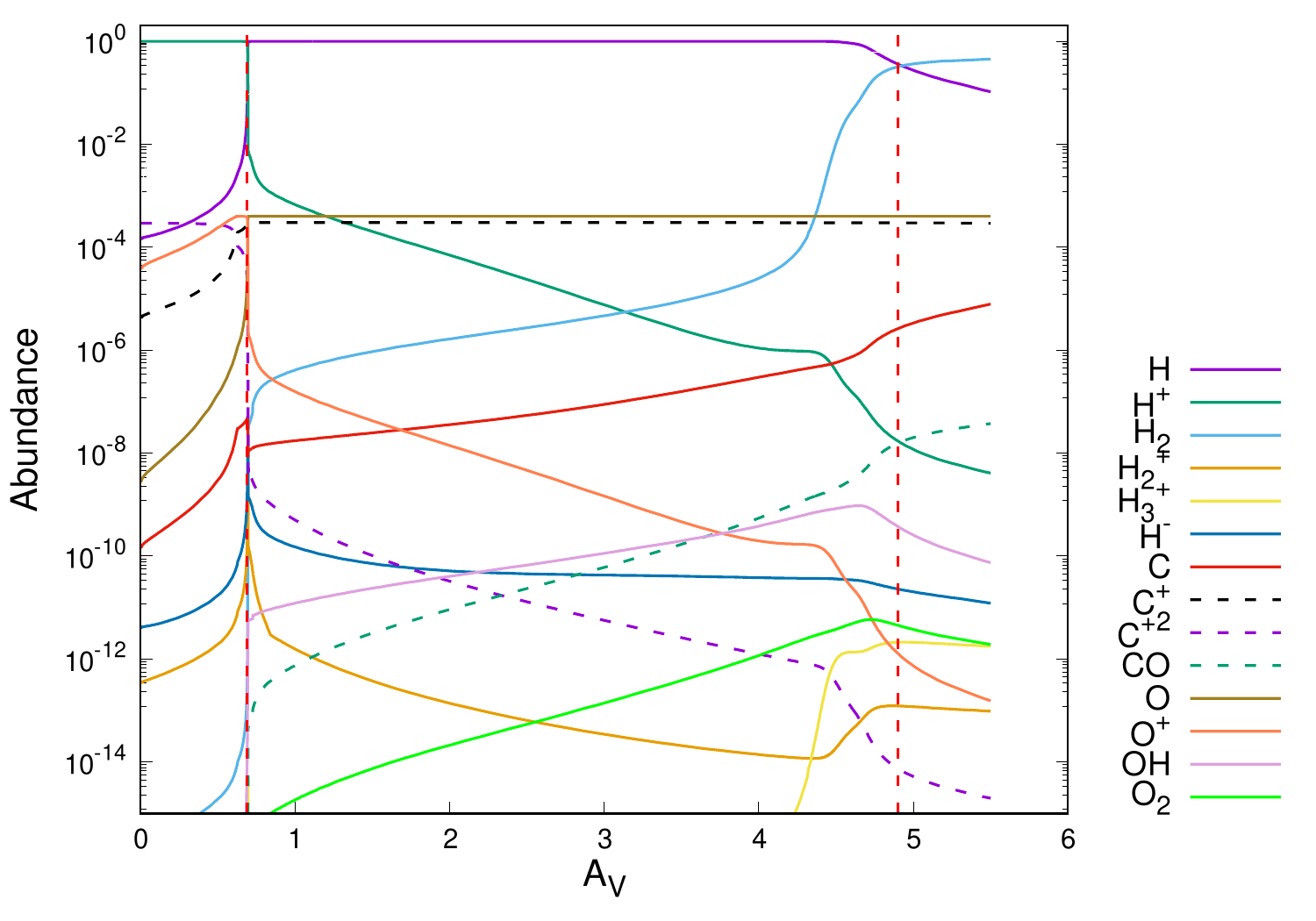}
    \caption{Upper panel: Total hydrogen number density, electron number density, and temperature variation. Lower panel: Abundance variation of various H, C, and O-related species with respect to total hydrogen nuclei in the H\,{\sc ii} to PDR obtained from the \textsc{Cloudy} code.
    \label{fig:pdr-phys}}
\end{figure}

\subsection{Grain surface reactions \label{sec:gr_reac}}
In CMMC code, we consider an extensive grain surface network. This network is primarily based on the network provided by \cite{ruau16}. It also considers the H-abstraction reactions proposed by \citet{bell14} and \cite{garr17}.
However, in \textsc{Cloudy}, we selectively include some surface reactions relevant to this study.

Apart from the various essential surface reactions considered in the CMMC code, the following are important grain surface reactions relevant to this work. These reactions are also included in the \textsc{Cloudy} network. It is important to note that the \textsc{Cloudy} network includes very limited grain surface chemistry, and therefore, we do not utilize this code in scenarios where molecular depletion is significant and surface chemistry is crucial.

\begin{center}
Si$_{\rm gr}$ + H$_{\rm gr}$ $\rightarrow$ SiH$_{\rm gr}$, \\
SiH$_{\rm gr}$ + H$_{\rm gr}$  $\rightarrow$ SiH${_2}_{\rm gr}$,\\
SiH${_2}_{\rm gr}$ + H$_{\rm gr}$ $\rightarrow$ SiH${_3}_{\rm gr}$,\\
{SiH$_3$}$_{\rm gr}$ + H$_{\rm gr}$ $\rightarrow$ SiH${_4}_{\rm gr}$,\\
Na$_{\rm gr}$ + H$_{\rm gr}$ $\rightarrow$ NaH$_{\rm gr}$,\\
Fe$_{\rm gr}$ + H$_{\rm gr}$ $\rightarrow$ FeH$_{\rm gr}$,\\
Mg$_{\rm gr}$ + H$_{\rm gr}$ $\rightarrow$ MgH$_{\rm gr}$,\\
MgH$_{\rm gr}$ + H$_{\rm gr}$ $\rightarrow$ MgH${_2}_{\rm gr}$,\\
K$_{\rm gr}$ + H$_{\rm gr}$ $\rightarrow$ KH$_{\rm gr}$,\\
Al$_{\rm gr}$ + H$_{\rm gr}$ $\rightarrow$ AlH$_{\rm gr}$,\\
Na$_{\rm gr}$ + Cl$_{\rm gr}$ $\rightarrow$ NaCl$_{\rm gr}$,\\
Na$_{\rm gr}$ + S$_{\rm gr}$ $\rightarrow$ NaS$_{\rm gr}$,\\
K$_{\rm gr}$ + Cl$_{\rm gr}$ $\rightarrow$ KCl$_{\rm gr}$,\\
NaH$_{\rm gr}$ + Cl$_{\rm gr}$ $\rightarrow$ NaCl$_{\rm gr}$+H${_{\rm gr}}$\footnote[2]{\rm \scriptsize it \ is \ not \ included \ in \ the \ \textsc{Cloudy} chemistry \ network.}\\
\end{center}

Moreover, we consider cosmic-ray induced desorption and photo-desorption of these species. Since the cosmic-ray induced desorption and photo-dissociation rates were given for some of the Si-related species, we consider a similar rate for the other metallic compounds considered in our network for simplicity.

\begin{table*}
\scriptsize \centering
\caption{Initial elemental abundances considered for the H\,{\sc ii}-PDR, diffuse clud, dark cloud, and hot core cloud models with respect to total hydrogen nuclei.
\label{tab:Initial}}
\centering
\begin{tabular}{|cc|cc|cc|cc|cc|}
\hline
\multicolumn{2}{|c|}{\bf PDR$^a$}&\multicolumn{2}{|c|}{\bf Diffuse cloud$^b$}&\multicolumn{2}{|c|}{\bf Dark cloud$^c$}&\multicolumn{2}{|c|}{\bf Hot core$^d$}\\
Species	& Abundances &Species	& Abundances&Species	& Abundances&Species	& Abundances \\
\hline
         &&&& H$_2$&$5.0 \times 10^{-1}$&H$_2$&$4.99 \times 10^{-1}$\\
         H&$1.0 \times 10^0$  &H&$1.0 \times 10^{0}$& H&$5.0 \times 10^{-5}$&H&$2.0 \times 10^{-3}$ \\
         He&$9.5 \times 10^{-2}$&He&$8.5 \times 10^{-2}$&He&$9.0 \times 10^{-2}$&He&$9.0 \times 10^{-2}$\\
         C&$3.0 \times 10^{-4}$&C$^+$&$2.7 \times 10^{-4}$&C$^+$&$1.4 \times 10^{-4}$&C$^+$&$1.4 \times 10^{-4}$\\
         N&$7.0 \times 10^{-5}$&N&$6.8 \times 10^{-5}$& N&$7.5 \times 10^{-5}$&N&$7.5 \times 10^{-5}$ \\
         O&$4.0 \times 10^{-4}$&O&$4.9 \times 10^{-4}$&O&$3.2 \times 10^{-4}$&O&$3.2 \times 10^{-4}$\\
         &&F&$3.6 \times 10^{-8}$& F&$2.0 \times 10^{-8}$&F&$2.0 \times 10^{-8}$\\        
         Na&$3.0 \times 10^{-7}$&Na$^+$&$1.7 \times 10^{-6}$&Na$^+$&$2.0 \times 10^{-9}$&Na$^+$&$2.0 \times 10^{-8}$\\
         Mg&$3.0 \times 10^{-6}$&Mg$^+$&$3.9 \times 10^{-5}$&Mg$^+$&$7.0 \times 10^{-9}$&Mg$^+$&$7.0 \times 10^{-9}$\\
         Al&$2.0 \times 10^{-7}$&Al$^+$&$2.3 \times 10^{-6}$&Al$^+$&$2.7 \times 10^{-10}$&Al$^+$&$2.7 \times 10^{-10}$\\
         Si&$4.0 \times 10^{-6}$&Si$^+$&$3.2 \times 10^{-5}$&Si$^+$&$8.0 \times 10^{-9}$&Si$^+$&$8.0 \times 10^{-9}$\\
         &&P$^+$&$2.6 \times 10^{-7}$&P$^+$&$3.0 \times 10^{-9}$&P$^+$&$3.0 \times 10^{-9}$ \\
         S&$1.0 \times 10^{-5}$&S$^+$&$1.3 \times 10^{-5}$& S$^+$&$1.5 \times 10^{-6}$&S$^+$&$8.0 \times 10^{-8}$ \\
         Cl&$1.0 \times 10^{-7}$&Cl$^+$&$3.2 \times 10^{-7}$&Cl$^+$&$4.0 \times 10^{-9}$&Cl$^+$&$4.0 \times 10^{-9}$\\
         K&$1.1 \times 10^{-8}$&K$^+$&$1.2 \times 10^{-7}$& K$^+$&$1.2 \times 10^{-10}$&K$^+$&$1.2 \times 10^{-10}$\\
         Fe&$3.0 \times 10^{-6}$&Fe$^+$&$3.2 \times 10^{-5}$&Fe$^+$&$3.0 \times 10^{-9}$&Fe$^+$&$3.0 \times 10^{-9}$\\         
\hline
\end{tabular}\\
\tablefoot{\footnotesize
$^a$ Initial abundances are considered from \cite{shaw22}, except Na, K, and Al.\\
{$^b$ Initial abundances are considered from \cite{chan20}, except for Al and K, which are set to standard solar abundances from \cite{aspl09}. Additionally, F is considered neutral.}\\
$^c$ Initial abundances are considered from \cite{mill22} except for Al and K.\\
$^d$ Initial abundances are considered from \cite{garr17} except for Al, K, and F.}
\end{table*}

\section{ Chemical models for various regions}
\label{sec:models}

We utilized \textsc{Cloudy} v22.01 to model the H\,{\sc ii}-PDR. Due to its ability to consider various microphysics, the \textsc{Cloudy} code is useful in studying chemical compositions of H\,{\sc ii}-PDR \citep{das20,Sil2024b,Sil2025}. Additionally, we employed the CMMC code to model the diffuse cloud, dark cloud, and hot core regions. For the gas-phase chemistry, \textsc{Cloudy} v22.01 was based on the UMIST-2012 database \citep{mcel13}, while the CMMC code used the UMIST-2022 gas-phase network \citep{mill22}. It's worth noting that \textsc{Cloudy} did not consider grain surface reactions extensively, whereas the CMMC code employed an extensive grain surface chemical network.

\subsection{H\,{\sc ii} region and PDR \label{sec:pdr}}
H\,{\sc ii} regions are created when the extreme ultraviolet (UV) radiation from a young massive star emits copious amounts of photons beyond the Lyman limit ($h\nu > 13.6$~eV) and ionizes and heats its surrounding molecular cloud.
The PDR is often associated with the H\,{\sc ii} region.
A PDR is a neutral atomic region where the column density of the cloud is insufficient to block the far-UV photon (6 eV $<h\nu < 13.6$ eV). These far-UV photons are the primary energy sources to glow this region in the infrared (IR) and regulate various chemical processes crucial for star formation in galaxies \citep{holl97}.
Atomic hydrogen is the most abundant here, but a little deeper inside the cloud, where the radiation field is attenuated, the formation of molecular hydrogen starts. Similarly, the transition of C$^+$ to CO around this region starts.
Here, we use the models by \citet{abel05,shaw22} to self-consistently model the thermal and chemical structure of the H\,{\sc ii} and PDR. We adopted the non-LTE CoStar stellar atmospheres with a temperature of the ionizing star ($T_\star)$  $ \sim 39\,600$~K, numbers of ionizing photons per second (Q(H)) $\sim 10^{49}$, and a photon flux of per unit area of cloud surface ($\Phi(H)$) $\sim 10$ cm$^{-2}$s$^{-1}$. The separation between the star and the cloud inner surface of $\sim 2.82 \times 10^{17}$ cm with a closed geometry is considered for this calculation.
A gas-phase abundance based on average abundances obtained in the Orion Nebula is used \citep{bald96,abel05}. We consider a Galactic cosmic-ray ionization rate having a mean H$^0$ ionization rate ($\zeta_H$) of $2.0 \times 10^{-16}$~s$^{-1}$.
Since few grain surface reactions were included in the \textsc{Cloudy} v22.01, we stopped our calculations at $A_V=5.5$~mag to avoid misleading results.
We use the initial elemental abundances
listed in the second column of  Table~\ref{tab:Initial}. These are the abundances for the H\,{\sc ii} region, such as the Orion Nebula
\citep{bald91,oste92,rubi93}.
Na, K, and Al were not considered in the model of \citet{shaw22} and \cite{abel05}.
We consider an initial elemental abundance of Na, K, and Al of $3.0 \times 10^{-7}$,  $1.1 \times 10^{-8}$, and $2.00 \times 10^{-7}$, respectively. The MRN size distribution \citep{math77} is used with $\rm{a_{min}=0.03}$~\micron and $\rm{a_{max}=0.25}$~\micron to produce the Orion extinction curve around $R_V=5.5$ \citep{bald96}. With this grain size distribution, we obtain $R_V=5.32$.

In Fig.~\ref{fig:pdr-phys}, the upper panel displays variations in number density (hydrogen and electron) and temperature with visual extinction, while the lower panel shows the variation in abundance for different H, C, and O species.
A closed geometry is considered with a central star surrounded by a cloud ring. The increase in visual extinction on the right side indicates that the left side is the illuminated face. 
{The two red dashed vertical lines in the figure indicate the PDR starting from the boundary of the H~{\sc{ii}} region (ionization front (IF) where half of the H$^+$ has recombined) up to the H/H$_2$ transition zone (dissociation front (DF) where molecular hydrogen number density reaches 50\% of the number density of total hydrogen nuclei).}

According to the data presented in the lower panel, at the beginning of the PDR, all the H$^+$ ions are transformed into atomic H, and O$^+$ is converted into atomic O. Carbon, on the other hand, remains largely in C$^+$ form. As we move towards the deeper parts of the cloud ($A_V>5$), the formation of H$_2$ and CO begins. Over time, H$_2$ and CO become some of the most abundant molecules inside the cloud.

Since the \textsc{Cloudy} code does not consider substantial surface chemistry, we restricted our simulation to $A_V=5.5$. The fate of these molecules in the UV-shielded region is discussed later using the CMMC code 
for the diffuse, dark, and hot core region in sections \ref{sec:DIFF}, \ref{sec:dark}, and \ref{sec:hc}, respectively.

\subsection{Diffuse cloud} \label{sec:DIFF}
We adopted standard solar initial elemental abundances as described by \citep{aspl09}. Initially, all hydrogen is considered in its atomic form. The species listed in the third column of Table \ref{tab:Initial} are regarded as ionized if their ionization potential is lower than that of hydrogen. 
A similar set of elemental abundances was also used by \cite{chan20} for the diffuse cloud. The only differences are the inclusion of ionized aluminum and potassium in the initial elemental abundance and the consideration of initial fluorine in its neutral form, given its high ionization potential compared to hydrogen.

In this study, we utilize the CMMC code and assume a total hydrogen nuclei number density ($n_H$) of { $500$ cm$^{-3}$, with an extinction value of 2} and a gas temperature of $\rm{T_{gas}}=40$ K.  Similar physical parameters were used by \citet{chan20} and \cite{sil21} in their diffuse cloud models.
Recent studies \citep{obol24,neuf24} have suggested that the gas densities in the foreground interstellar clouds responsible for the observed C$_2$ absorption are a factor of 4 to 7 times smaller than previously inferred. Based on this finding, the consideration of $n_H=500$ cm$^{-3}$ for the number densities of local diffuse clouds may be an overestimate. Therefore, we examine an alternative case for the diffuse cloud with $n_H = 50$ cm$^{-3}$.
Dust temperature is derived from the value of $A_V$ using the empirical relation developed by \citet{garr11}. For $A_V = 2$, the derived dust temperature is approximately 17 K. 
We use a cosmic ray ionization rate ($\zeta_{\rm CR}$) $1.7 \times 10^{-16}$ s$^{-1}$ as was derived by \cite{indr12}.
Nonthermal desorption is crucial for exchanging chemical components between gas and grain. We consider the reactive desorption of surface species, leading to a single product \citep{garr07}. 
The energy released by the reaction is calculated from the difference between the enthalpies of the formation of products and reactants. Our calculated enthalpies of formation (see Table~\ref{tab:enthalpy}) are used in this estimation. The photodesorption yields of pure CO ice were determined to be $3 \times 10^{-3}$ molecules per UV ($7-10.5$~eV) photon at 15~K in an experimental study by \citet{ober07}. A significantly higher photodesorption yield of CO molecules, exceeding the former by a factor of ten, was noted by \citet{muno10}. In our model, we adopt a photodesorption yield of $3 \times 10^{-3}$ molecules~photon$^{-1}$ for all species. The photodesorption rate is estimated using the empirical relation in \citet{oberg09}.

\subsection{Dark cloud \label{sec:dark}}
For the dark cloud model, we consider $n_H= 2 \times 10^4$ cm$^{-3}$, a gas temperature of {10 K}, a dust temperature of 10 K, and $A_V=10$. Following \cite{mill22}, an initial low metallic depleted elemental abundance shown in the sixth column of Table~\ref{tab:Initial} is considered for the modeling.
\citet{aspl09} obtained a solar abundance of Al and K as $2.82 \times 10^{-6}$ and $1.07 \times 10^{-7}$, respectively.
Elements like Al suffer from heavy depletion. In diffuse clouds, it is suggested that 90-99$\%$ of Al could be depleted into grains \citep{turn91}.
Considering the same depletion factor for Al as Fe ($\sim$ 10\,540 times lower than the solar abundance), we start with an initial Al-abundance of $2.67 \times 10^{-10}$.
Similarly, for the initial elemental abundance of K, we use the same depletion factor as it was considered for Na ($\sim$ 870 times lower than the solar abundance). So, here we start with an initial K-abundance of $1.23 \times 10^{-10}$.
A standard H$_2$ cosmic-ray ionization rate of $1.3 \times 10^{-17}$ s$^{-1}$ is considered. We use the CMMC code for this case and continue our simulation for $10^6$ years. The other chemical parameters are the same as in Sect.~\ref{sec:DIFF}.

\subsection{Hot core \label{sec:hc}}
Here, we consider CMMC code to model the hot core region. 
The construction of a toy model for this region is delineated into four distinct phases. The first phase is characterized by a quasi-static collapse, wherein the hydrogen number density of the cloud experiences a linear increase from \(3 \times 10^3 \, \text{cm}^{-3}\) at \(t = 0\) to \(2 \times 10^4 \, \text{cm}^{-3}\) at \(t = 10^6\) years. Concurrently, a linear rise in the extinction value (\(A_V\)) is assumed, progressing from 2 to 10 during this timeframe. The gas temperature (\(T_{gas}\)) is maintained at a steady value of 10 K, while the dust temperature (\(T_{dust}\)) is varied according to the relation presented by \citet{garr11,zucc01}. A minimum dust temperature is set at 10 K for this phase.

In the subsequent phase (free-fall collapse phase), the density increases from \(2 \times 10^4 \, \text{cm}^{-3}\) to \(10^7 \, \text{cm}^{-3}\) throughout approximately \(3.58 \times 10^5\) years following the modified free-fall collapse described in \cite{rawl92}. During this stage, it is assumed that the gas and dust temperatures remain well coupled, sustaining a constant temperature of 10 K. The visual extinction parameter increases following the relationship:
\[
A_V = A_{V_0} \left(\frac{n_H}{n_{H_0}}\right)^{2/3},
\]
where \(n_{H_0}\) and \(A_{V_0}\) correspond to the initial density $\sim$ \(3 \times 10^3\, \text{cm}^{-3}\)) and visual extinction $\sim 2$ established at the onset of the quasi-static collapse phase.

The third phase, the warm-up stage, maintains the density and visual extinction at their maximum levels attained at the end of the free-fall phase, specifically at \(t = 1.358 \times 10^6\) years. During this interval, the gas and dust temperatures are permitted to rise linearly, reaching up to 400 K over \(5 \times 10^4\) years, ending at \(t = 1.363 \times 10^6\) years. The final stage extends for an additional \(10^5\) years, during which the density, temperature, and visual extinction parameters are upheld at their peak values established during the warm-up stage.

\begin{figure}
    \centering
    \includegraphics[width=\linewidth]{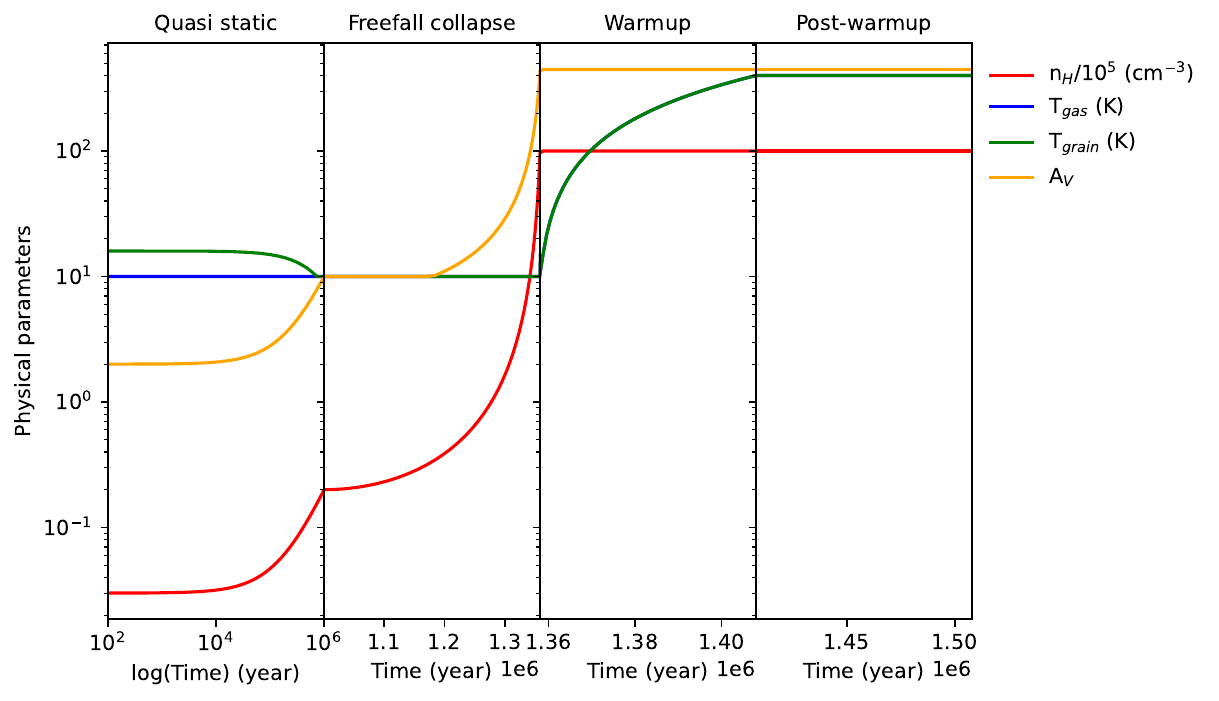}
    \caption{Adopted physical condition in the hot core model.
    \label{fig:phys-HC}}
\end{figure}

We use the initial elemental abundances presented in the eighth column of Table~\ref{tab:Initial} for our modeling. These initial abundances are the same as those considered in \citet{garr17,sriv22,rama24}, except for the initial elemental abundance of Al and K. For Al, we used the same initial elemental abundance ($\sim 2.67 \times 10^{-10}$) as in Sect.~\ref{sec:dark}, following the same depletion factor of Fe from its solar abundance. Although \citet{garr17} and \cite{sriv22} used an order of magnitude higher initial elemental abundance of Na ($2 \times 10^{-8}$) than \citet[][$2 \times 10^{-9}$]{wake17}, we adopt an abundance of K of $1.23 \times 10^{-10}$ for the hot core model. We use a similar abundance of F (approximately $2 \times 10^{-8}$) as considered in our dark cloud model. Figure~\ref{fig:phys-HC} shows the physical parameters and time scales used in our modeling.

\section{Results obtained and astrophysical implications \label{sec:astro}}
Extensive searches for metallic compounds in diffuse and molecular clouds have been conducted. However, many of these searches have been unsuccessful due to the depletion of refractory elements. It is commonly believed that shocks primarily release refractory elements, which might be present in the molecular cloud, albeit at a low level. In molecular clouds, detailed metal-related chemistry is usually not considered in the chemical network. However, the UMIST-2022 database has recently updated many such reactions. We have added some more reactions noted in Table A.1 (available on Zenodo) in the gas-phase network to provide a complete list of reactions that could be included. 
Implementing these additional pathways discussed in Sect.~\ref{sec:chem-route} could significantly impact the abundances of metal-related species, which may subsequently alter the abundances of their related species.

\begin{figure}
    \centering
    \includegraphics[width=\linewidth]{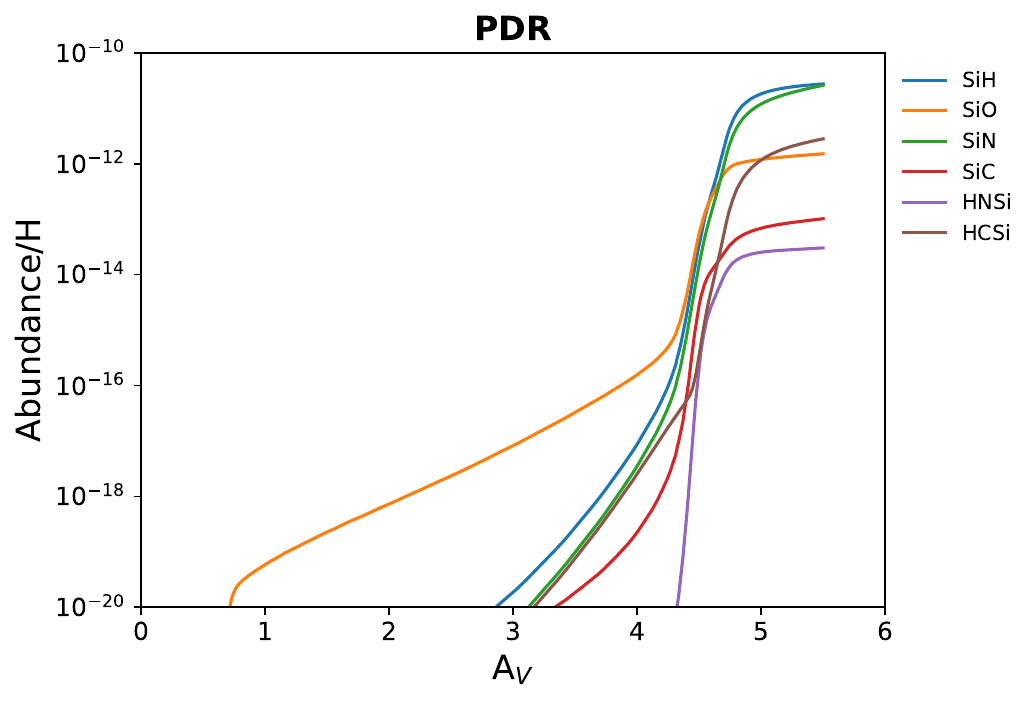}
    \caption{The abundances of some notable metal-bearing species in the PDRs.}
    \label{fig:PDR-abun}
\end{figure}

\begin{figure}
    \centering
    \includegraphics[width=\linewidth]{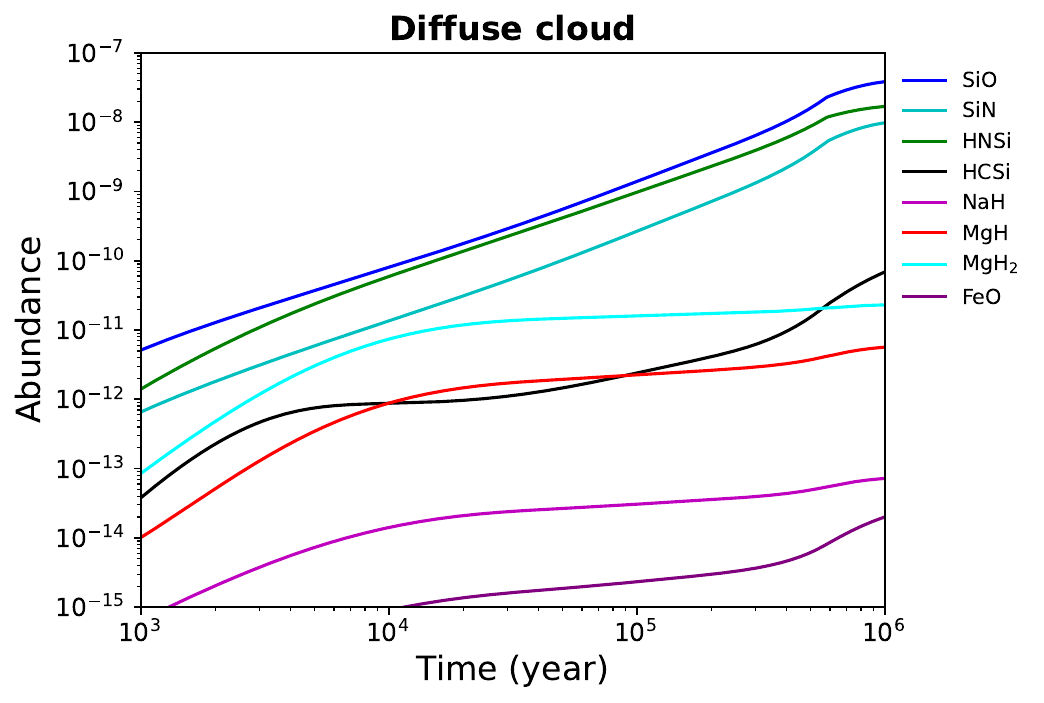}
    \caption{Time evolution of abundances of some notable metal-bearing species in the diffuse clouds (n$_H=50$ cm$^{-3}$).}
    \label{fig:DIFF-abun}
\end{figure}

\begin{figure}
    \centering
    \includegraphics[width=\linewidth]{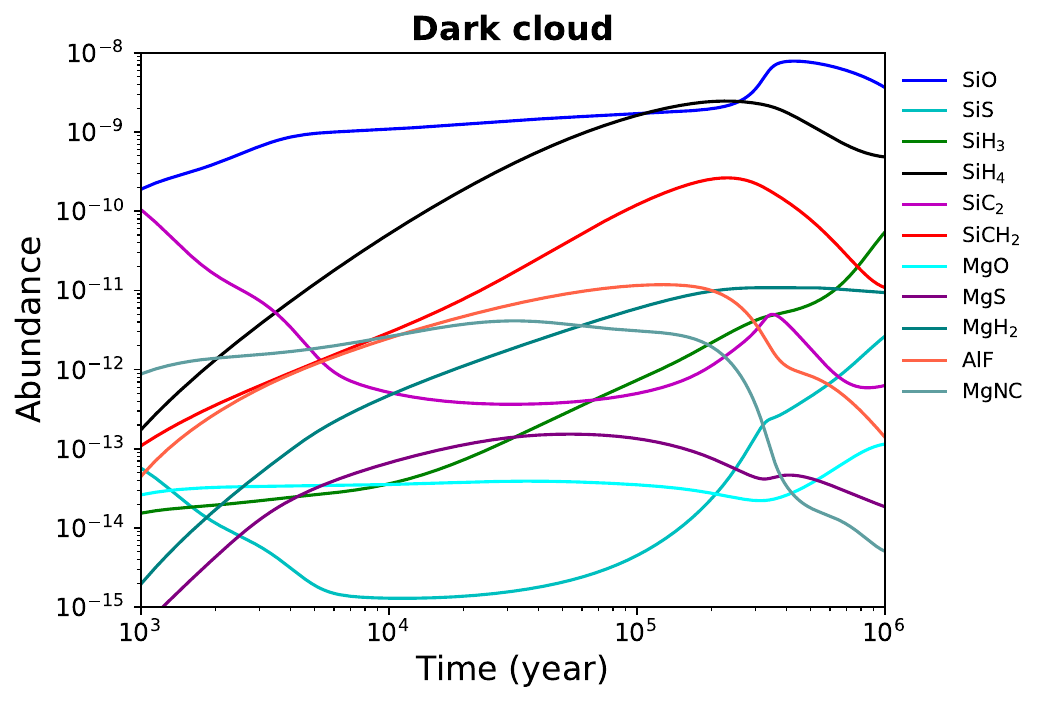}
    \caption{Time evolution of abundances of some notable metal-bearing species in the dark clouds.}
    \label{fig:Dense-abun}
\end{figure}

\begin{figure}
    \centering
    \includegraphics[width=\linewidth]{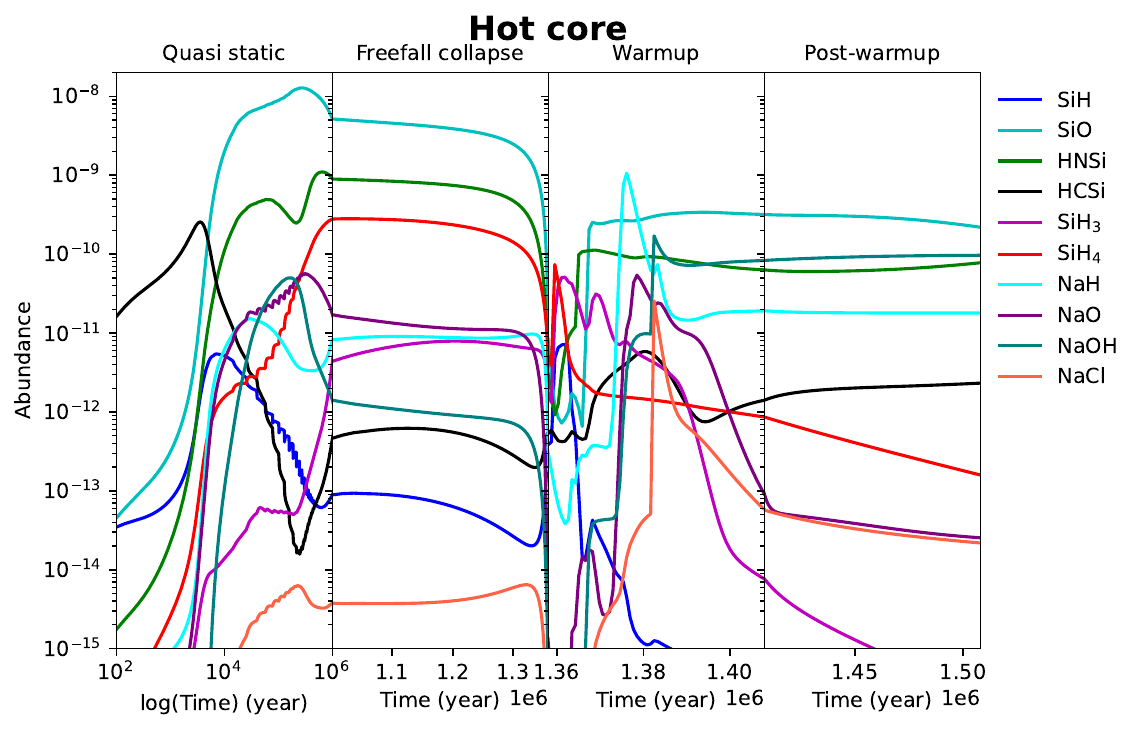}
    \includegraphics[width=\linewidth]{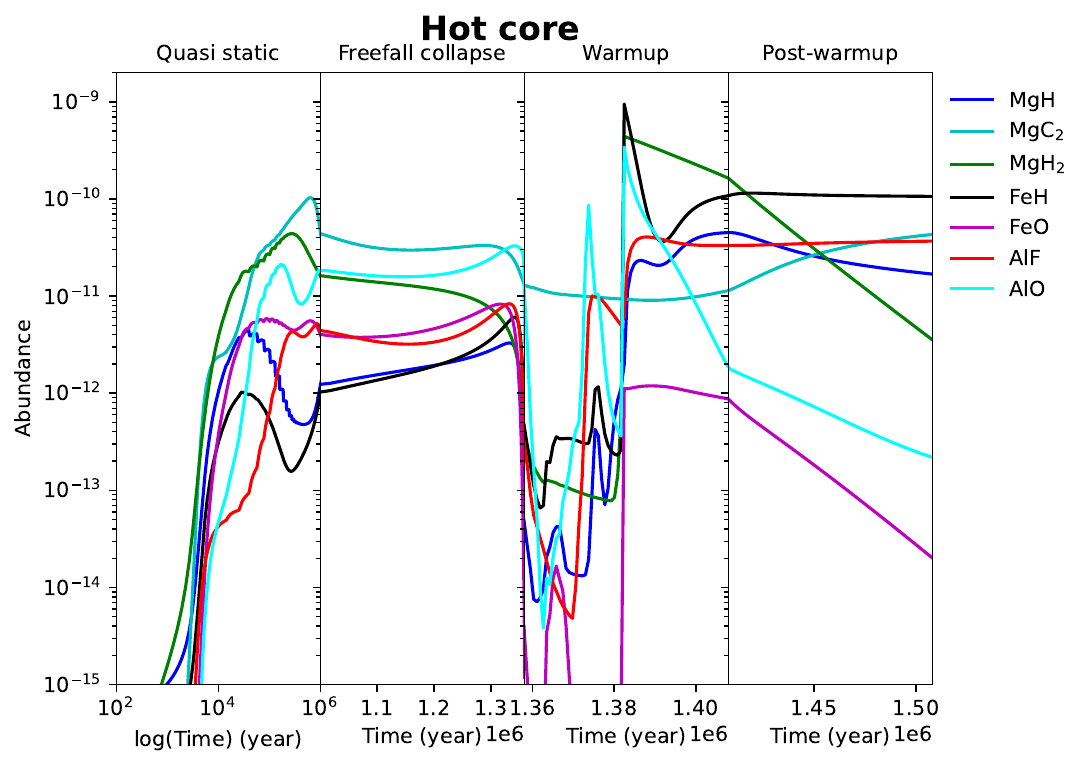}
    \caption{Time evolution of abundances of some notable metal-bearing species in the hot cores.}
    \label{fig:HC-abun}
\end{figure}

The modeled abundances of specific metallic species, those that may be observable and exceed levels of \(10^{-14}\) relative to H nuclei, are presented for PDRs in Table \ref{table:pdr}. For the diffuse, dense, and hot core models, the obtained abundances with our network and with the UMIST-2022 network are shown in Tables \ref{table:diff}, \ref{table:dense}, and \ref{table:HC}, respectively, with levels exceeding \(10^{-12}\) relative to H$_2$.
In these tables, we have also included references for earlier observations, along with their estimated abundances or upper limits for these regions. According to our modeling results, the abundance profiles of notable metal-bearing species are shown in Figures \ref{fig:PDR-abun}, \ref{fig:DIFF-abun}, \ref{fig:Dense-abun}, and \ref{fig:HC-abun} for the PDRs, diffuse clouds, dark clouds, and hot cores, respectively.

In general, it is noticed that the key reactions that influence the abundances of overall metal-bearing species are:
\(\rm M^+ + H_2 \rightarrow MH_2^+ + h \nu\);
\({\rm MH + O \rightarrow MO + H}\); \(\rm{M_1^+ + M_2H \rightarrow M_1H + M_2^+}\), where M, M$_1$, M$_2$ are metals.  
In this section, we present our compelling modeling results along with a series of observational attempts specifically conducted in the modeled regions.

\subsection{Photon-dominated regions}
Not many metal-bearing molecules have been found in PDRs because high UV radiation prevents the formation of stable molecular bonds. Any metal-bearing molecules that do form are quickly broken apart by UV photons. Table \ref{table:pdr} summarizes the abundances from our PDR model, including only those with abundances greater than \(10^{-14}\) relative to hydrogen nuclei. 

\subsubsection{Metal-bearing species in PDRs: Detection attempts and model predictions}
\cite{schi01b} utilized the IRAM 30m telescope to observe SiO transitions in PDRs such as the Orion Bar and S 140. SiO was detected with abundances around $10^{-11}$ relative to H$_2$. Moreover, they also put limits of $(2-4) \times 10^{-11}$ towards several positions in NGC 2023 and NGC 7023. The findings suggest that while silicon is heavily depleted onto dust grains, a fraction returns to the gas phase near ionization fronts, possibly due to UV-driven processes or mild shocks. Our PDR model finds an abundance of $3.0 \times 10^{-12}$ relative to H$_2$. 

\subsubsection{Other noteworthy species in PDRs}
 Table \ref{table:pdr} indicates that among the metal-bearing molecules, only some silicon-containing species are abundant in PDRs, with abundances exceeding \(10^{-12}\). Specifically, SiH and SiC have abundances of approximately \(10^{-11}\), while HCSi is roughly \(10^{-12}\) relative to hydrogen. Moreover, as highlighted in Table \ref{tab:dipole}, it is noteworthy that both SiH and HCSi exhibit very weak dipole moments. This characteristic presents certain challenges when attempting to observe them through pure rotational transitions in PDRs.

\subsection{Diffuse clouds}
As with the PDRs, a limited number of metal-bearing molecules have been reported in the diffuse region due to the low densities and intense UV radiation. However, some attempts were made to estimate the upper limits for metal-bearing molecules in diffuse clouds. 

\subsubsection{Metal-bearing species in Diffuse clouds: Detection attempts and model predictions}

\cite{czar87} searched for NaH and MgH in diffuse clouds using UV absorption spectroscopy but did not detect either molecule. They set upper limits for NaH and MgH with column densities of NaH  $\leq 1.3 \times 10^{10}$ cm$^{-2}$ and MgH $\leq 1.9 \times 10^9$ cm$^{-2}$ along the sightline to $\zeta$ Ophiuchi.
NaH and MgH in our network primarily formed by grain surface reactions (\( \text{Na$_{\rm gr}$} + \text{H$_{\rm gr}$} \rightarrow \text{NaH$_{\rm gr}$} \), \(\text{Mg$_{\rm gr}$} + \text{H$_{\rm gr}$} \rightarrow \text{MgH$_{\rm gr}$} )\).
NaH has a large dipole moment; however, based on our model, its abundance is approximately $\sim 10^{-14}-10^{-13}$, which suggests it is likely not observable. For MgH, we observe that our estimated abundance falls within the range of approximately \(10^{-12}\) to \(10^{-10}\), indicating that it may be within the detectable limits.

\cite{luca00} estimated a SiO abundance of $(2-20) \times 10^{-11}$ for diffuse and translucent clouds that lie towards the extragalactic continuum sources.
One transition of FeO was identified towards the ultra compact H\,{\sc ii} region associated with SgrB2(M) in absorption \citep{walm02,furu03}.
\citet{walm02} derived an abundance ratio of $\sim 0.002$ for FeO/SiO with FeO abundance of $\sim 3 \times 10^{-11}$ with respect to H$_2$. They estimated that the density of the absorbing gas is at a relatively low density and high temperature ($\sim 500$~K). They suggested that the lower abundance of FeO would indicate that iron may be liberated by shocks, but it has yet to be processed in molecular form. \citet{schi03} reported the detection of SiN  having a very high abundance $\sim 10^{-7}$ relative to H nuclei toward SgrB2(M) in absorption.
Moreover, \citet{schi97} estimated that 10\% of the Si would be released by the shock and could contribute to the Si-related chemistry in the gas phase. They estimated a low density of $\rm{n_{H2}=10^3}$ cm$^{-3}$ and a high temperature of $\sim 500$ K in the cloud, which may explain this feature. Compared to SiO, \citet{schi03} estimated that the column density of SiN is 20 to 30 times lower. 

While our diffuse cloud model may not be directly applicable under these conditions, it is still valuable to compare the abundances of FeO, SiN, and SiO derived from our model. Notably, as shown in Table \ref{table:diff}, we found FeO/SiO and SiO/SiN ratios of 0.001 and 10, respectively, for a hydrogen density (n$_H$) of 500 cm$^{-3}$. The table illustrates that we observed a significantly increased abundance of FeO when additional pathways were considered. In this case, the formation of FeO is primarily driven by gas-phase reactions, particularly through the exothermic reaction: $\rm{FeH+O\rightarrow FeO+ H}$.

\subsubsection{Other noteworthy species in diffuse clouds}
Among the other silicate-bearing species, our study indicates that HNSi, SiC, and HCSi are abundant in diffuse clouds. Like SiN, the formation of HNSi appears through the dissociative recombination of $\rm{SiNH_2}^+$. Its formation is also significant by $\rm{N+ SiH_2 \rightarrow HNSi + H}$ and $\rm{N+ SiH_3 \rightarrow HNSi + H_2}$.
HNSi can react with the major ions like ${\rm H_3}^+$, ${\rm H_3O^+}$, and HCO$^+$ to produce $\rm{SiNH_2}^+$. SiN is destroyed by oxygen ($\rm{SiN + O}$), which doesn't apply to HNSi, possibly explaining why HNSi abundance is higher than that of SiN in our model. 
Interestingly, HNSi has not yet been detected in any astronomical observations. \citet{pari00} attributed this discrepancy to the weak dipole moment of HNSi (0.25 Debye, see Table \ref{tab:dipole}) or the possibility that dissociative recombination of $\rm{SiNH_2}^+$ might not produce HNSi.
HCSi has not yet been detected in space. Our diffuse cloud model, predicts a high abundance of HCSi and SiC. As indicated in Table \ref{tab:dipole}, HCSi exhibits a modest dipole moment of 0.07 Debye. This small value could potentially account for its current challenges in detection.

Our model indicates a significant abundance of MgH$_2$ at approximately $10^{-11}-10^{-9}$. Despite having such a high abundance, its non-detection in the diffuse region may be attributed to its low dipole moment (0.18 Debye, see Table \ref{tab:dipole}). 
Table~\ref{table:diff} illustrates an abnormally high abundance (on the order of a few times \(10^{-7} -10^{-5}\)) of \ce{MgC2}, \ce{MgC2H}, \ce{MgC6H}, and \ce{MgC8H} in our diffuse cloud with n$_H$=500 cm$^{-3}$. However, with $n_H=50$ cm$^{-3}$, we obtain a negligible abundance of these species. 
With the $n_H=500$ cm$^{-3}$, the formation of \ce{MgC2}, \ce{MgC2H}, \ce{MgC6H}, and \ce{MgC8H} can be processed through a series of dissociative recombination reactions:
$\rm{MgC_xH_2^+ + e^- \rightarrow MgC_xH/MgC_2 + C_yH/C_yH_2}$ 
(x=2, 4, 6, 8 and y = 0, 2, 4, 6).
The formation of MgC$_2$ is also significantly influenced by the dissociative recombination of MgC$_2$H$^+$. During the initial phase, C$_2$H$_2$ is abundant; however, over time, larger hydrocarbons such as C$_6$H$_2$ and C$_8$H$_2$ become more prevalent compared to smaller hydrocarbons like C$_2$H$_2$ and C$_4$H$_2$. The primary production of MgC$_2$, MgC$_2$H, MgC$_6$H, MgC$_8$H mainly results from larger hydrocarbons, specifically ${\rm MgC_8H_2}^+$ and ${\rm MgC_6H_2}^+$. These larger hydrocarbons associate with Mg$^+$ more quickly than the smaller ones, leading to an increased formation of ${\rm MgC_8H_2}^+$ and ${\rm MgC_6H_2}^+$ compared to ${\rm MgC_2H_2}^+$ and ${\rm MgC_4H_2}^+$.

\subsection{Dark clouds}
Metal-bearing molecules are rarely observed in dark clouds because metals are largely depleted onto dust grains, and the cold, dense conditions inhibit their desorption into the gas phase. Additionally, the emission lines of these molecules are often too weak to be detected with current sensitivity limits. However, several attempts have been made over the last few decades to observe these species in this region and estimate their abundances or the upper limits of observation. 

\subsubsection{Metal-bearing species in Dark clouds: Detection attempts and model predictions}
\cite{turn05} conducted a sensitive search for MgNC and AlNC in dark cloud core, TMC-1, with the Green Bank Telescope, and reported an upper limit of $1.2 \times 10^{-11}$ for MgNC. We did not consider the formation of AlNC in our network. Our dark cloud model with new pathways obtained a negligible abundance ($\sim 10^{-14}$) of MgNC. We have seen a significant impact on the MgNC abundance when its destruction by H, O, and C$_2$H as suggested by \cite{turn05} is considered.

\cite{ziur89,ziur91} estimated an upper limit for SiO and SiS abundance to be $<2.4 \times 10^{-12}$ and $<5.9 \times 10^{-11}$, respectively, in the TMC-1. Our dark cloud model obtained a SiO and SiS abundance of $3.6 \times 10^{-9}$ and $2.6 \times 10^{-12}$, respectively. The abundance of SiO we obtained is notably higher than the upper limit established by \cite{ziur89}. It is to be noted that the new pathways have minimal impact on Si-bearing species. Thus, such a high SiO abundance is also obtained with the UMIST-2022 network as well.

Recently, several metal-bearing species have been identified in the quiescent molecular cloud G+0.693-0.27, located at the Central Molecular Zone of the Milky Way, or at least their upper abundance limits have been estimated \citep{mass23,mont24}. Our dark cloud model is more applicable for a source like TMC-1, which is cold (around 10 K) and well-shielded. This environment facilitates slow chemical processes dominated by grain surfaces. In contrast, G+0.693-0.027 is warmer (approximately $50-100$~K) and subject to comparatively intense radiation and shocks, which results in active gas-phase chemistry.

\subsubsection{Other noteworthy species in Dark clouds}
Recently, the detection of rotational lines of AlF ($\nu=0$, J = 7-6) toward the S-type AGB star W Aql was reported by \cite{dani21}. The first identification of AlF in M-type AGB stars was also provided by \cite{sabe22}. However, AlF has not yet been identified in the dark cloud environment.
The formation of AlF is included in the UMIST-2022 database. Table A.1 (available on Zenodo) presents some newly accounted ion-neutral destruction reactions for AlF. Without these additional pathways, the calculated abundance of AlF is \(2.7 \times 10^{-10}\), raising questions about why it has not been detected in the dark cloud. When considering newly accounted reactions, its abundance significantly drops to \(1.4 \times 10^{-13}\), a level that could explain its non-detection. 

SiH$_4$ is a non-polar spherical top molecule with a negligible rotationally induced dipole moment. \citet{gold84} identified Silane by its rotation-vibration spectrum.
The pure rotational transitions of SiH$_4$ would be extremely weak in the radio domain.
\citet{tatj15} studied the IR absorption spectra of SiH$_4$ at 11 K. With the nitrogen matrix, they obtained a triplet in the stretching (2206.4 cm$^{-1}$, 2203.0 cm$^{-1}$, and 2196.5 cm$^{-1}$) and a doublet in the bending region (910.3 cm$^{-1}$ and 910.2 cm$^{-1}$). Our modeling results suggest that in the dark cloud region, we have ample production of SiH$_4$ in the ice phase.
\cite{tenn16}(ExoMol) showed that SiH$_4$ has its largest cross-section in the mid-IR $\sim 4.5$~\micron, which could be accessible with the NIRSpec and PRISM instrument onboard the James Webb Space Telescope (JWST).
The bending doublet $\sim 10.98$~\micron could also fall under the accessibility of JWST. 
A search for SiH$_3$ was carried out at the mid-IR domain by \citet{cern17}, but it was unsuccessful.

According to our dark-cloud model, the abundance of SiCH$_2$ is estimated to be \(1.1 \times 10^{-11}\).
SiCH$_2$ has not yet been detected in space, primarily due to its low dipole moment (0.18 Debye, see Table \ref{tab:dipole}), which makes it challenging to identify in gas phase.  
We obtained a significant abundance ($9.4 \times 10^{-12}$) of MgH$_2$ from our dark cloud model. However, MgH$_2$ has a linear, symmetric structure with a zero permanent dipole moment, preventing it from emitting rotational transitions. As a result, it cannot be detected using radio astronomy techniques that rely on dipole-induced rotational spectra. The ice phase abundance of SiCH$_2$ and MgH$_2$ is approximately \( \sim 1.1 \times 10^{-10} \) and \( 7 \times 10^{-9} \) respectively. Given these abundances, the JWST may have the capability to detect these compounds through their IR vibrational transitions, which could provide valuable insights into their presence in various astrophysical environments.
\subsection{Hot cores}
Some metal-bearing molecules are detected in hot cores, as high temperatures and shocks may release metals from dust into the gas phase, enabling complex metal chemistry.

\subsubsection{Metal-bearing species in Hot cores: Detection attempts and model predictions}
SiO is ubiquitous in space. Generally, it is a good indicator of a shocked region. It is one of the first molecules discovered with radio astronomy \citep{wils71}. \citet{ziur91} estimated a SiO abundance of $\sim 1.2 \times 10^{-9}$ in Orion KL. Our hot core model achieved a peak abundance of SiO $\sim 3.4 \times 10^{-10}$ during the warm-up and post-warm-up stages.
\citet{schi01} tentatively identified SiH in Orion KL having fractional abundance of $\sim 3 \times 10^{-9}$. However, our hot core model underproduced SiH, having a peak gas-phase abundance of  $7.2 \times 10^{-12}$.
The observed underproduction of SiH and SiO in our model might be associated with the absence of shocks in the hot core model.

\citet{turn91} estimated an upper limit of $1.6 \times 10^{-9}$ for NaO in Orion KL. 
UMIST-2022 network did not consider the formation of NaO. The new pathways included in our network depict a peak gas-phase abundance of $5.4 \times 10^{-11}$ for NaO during the warm-up and post-warm-up stage of our hot core model. 
NaO in our model is mainly formed by the exothermic (having exothermicity of 8632~K) reaction: $\rm{NaH+O \rightarrow NaO + H}$.

Recently, NaCl has been detected in a massive protostellar system \citep{Gins19,wrig20}.
Our calculated binding energy indicates a high binding energy for NaCl (i.e., 15\,019~K; see Table~\ref{tab:be}). Interestingly, our hot core model obtains a reasonable abundance of NaCl (having a peak abundance of $2.6 \times 10^{-11}$). The formation of NaCl in our hot core model results from the surface reaction between NaH and Cl during the warm-up stage.

\subsubsection{Other noteworthy species in hot cores}
Though NaH has a significant dipole moment, it is yet to be identified in the hot core region because the warm and dense areas serve as sources for the local continuum, necessitating excitation above the background continuum to observe them in the submillimeter regime \citep{bern85}. 
In our hot core model, we observe a favorable formation of NaH and MgH. 
In the cold phase, the formation of these compounds is primarily driven by pathways on grain surfaces. However, during the warm-up and post-warm-up stages, the dissociative recombination of NaH${_2}^+$ and MgH${_2}^+$ plays a significant role. Additionally, certain exothermic channels of the reaction \( \text{M$_1$}^+ + \text{M$_2$H} \rightarrow \text{M$_1$H} + \text{M$_2$}^+ \) also make notable contributions to this process. Among the other hydrogen-bearing metals in hot cores, we have a notable abundance of MgH$_2$ (peak abundance $\sim 4.4 \times 10^{-10}$) and FeH ($\sim 9.5 \times 10^{-10}$). 

Like diffuse and dark clouds, HNSi also appears to be abundant in hot cores, with an estimated abundance of approximately \( \sim 10^{-10} \). Among the carbon-bearing magnesium compounds, MgC$_2$ and MgC$_2$H are found to be present at abundances of \( 4.3 \times 10^{-11} \) and \( 1.9 \times 10^{-12} \), respectively. The formation of MgC$_2$ in these phases is primarily driven by the dissociative recombination reaction of MgC$_2$H$^+$. Additionally, the formation of MgC$_2$H is mainly influenced by the dissociative recombination of MgC$_n$H$_2^+$, where \( n \) can be 2, 6, or 8.

Table \ref{table:HC} shows a significant increase in the abundance of FeO ($\sim 10^{-12}$) compared to when only the UMIST-2022 network is used. Additionally, we found that other notable species, such as AlO and AlF, are also abundant. The primary formation pathway identified for AlF is dominated by gas-phase reactions. ${\rm HF + Al \rightarrow AlF +H}$ and ${\rm AlO+HF \rightarrow AlF +OH}$. It is important to note that the abundance of AlO and FeO declines steadily after achieving a peak around $80-90$~K. During the warm-up stage, another oxygen-bearing species, likely NaOH, maintains a steady abundance of approximately \(10^{-10}\), making it an important species for future studies. Its formation in warm-up and post-warm-up stages is primarily influenced by the dissociative recombination of \({\rm NaH_2O}^+\).

\section{Conclusions \label{sec:con}}
We used various state-of-the-art modeling to explore the fate of metal-bearing species in some astrophysical environments. Our calculations with a comprehensive reaction network containing the metal-related species suggest revisiting many unsuccessful searches conducted in the 1970s and 1980s. In the following, we highlight the main results.

\begin{itemize}

\item The ionization process of a molecular cloud heavily relies on the depletion of metals, and it is crucial to have a realistic estimate of their binding energies. Quantum chemical calculations can be employed as a reliable alternative to laboratory measurements to make this estimation \citep{Sil2024a}. In dense media, water ice would be the most abundant surface species. In this study, we estimated the binding energy of some metals and their related species by considering water ice as the substrate. Based on their obtained binding energy values, they follow Fe, Si, Al, Na, K, and Mg in descending order.
We used the water substrate to provide the binding energy for several metallic species, which can benefit astrochemical modeling.
We found that the binding energies of some of the studied metals were significantly different from those used in earlier studies. For instance, we obtained much lower binding energies for Na and Mg (2408 K and 860 K, respectively), which were $5-6$ times lower than the earlier estimations. However, the obtained Fe's binding energy (16\,677~K) is about 5 times higher than previously known values. Our calculated binding energy for Si (7187~K) is consistent with the previously available calculations. Binding energy values for K and Al were not available, so we calculated them for the first time and obtained 2133 K and 4381 K, respectively. Another noteworthy is the binding energy obtained for SiH$_4$ (842~K), which is comparatively lower than that used in the past.
We noticed that MgNC, KH, and KCl favor chemisorption with the water surface, causing a change in molecular geometry. \\

\item  Ion-molecular reactions play a crucial role in interstellar environments, occurring efficiently even at low temperatures and densities, significantly affecting the chemical composition of the ISM. The dipole moments of some of the metallic species considered here were unavailable; therefore, quantum chemical calculations were employed to ascertain the dipole moments for these species (see Table \ref{tab:dipole}), which were subsequently integrated into the chemical model. \\

\item  A list of gas-phase reactions, as presented in Table~A.1 (available on Zenodo), was examined. The reaction enthalpies for several of these reactions were calculated using quantum chemical methods. Only those reactions that demonstrated exothermicity were incorporated into our network.\\

\item We have made a significant effort to compile a list of reactions that should be included in astrochemical models, particularly when dealing with molecular clouds. During our study, we found important sets of reactions that could affect model abundance estimation.

a) One such notable reaction is ${\rm M^+ + H_2 \rightarrow {MH_2}^+ + h}\nu$, where M=Na, Fe, Mg, Al, Si, and K, that could trigger the formation of ${\rm MH_2}^+$ and related chemistry. Usually, M=Si is considered in the network, but the others are neglected.

b) The reaction, ${\rm MH + O \rightarrow MO +H}$ (M=Na, Mg, Al, and Fe), proposed by \cite{turn91} is very important in dense regions for the production of metallic oxides.

c) $\rm{M_1^+ + M_2H \rightarrow M_1H + M_2^+}$ (where $\rm{M_1 \neq M_2}$, M$_1$,  M$_2=$ Na, Mg, Si, Al, K, Fe) is also found to be important.
\\

\item The destruction of MgNC by H, O, and C$_2$H was proposed by \cite{turn05}. In this study, we also consider similar destruction of MgCN, HMgNC, and MgC$_3$N, but only if their reaction enthalpy is exothermic (see Table~A.1, available on Zenodo). It is observed that the inclusion of these destruction processes significantly affects the abundances of these species.\\

\item In our study of the dense region, we observed that a significant amount of silicon gets trapped in the ice phase as SiH$_4$. Our findings indicate that SiH$_4$ has a much lower binding energy (842~K), so it may be released in the gas phase at significantly lower temperatures than previously assumed ($\sim 4500$~K according to the OSU database). According to our hot core model, we could have a peak gas-phase SiH$_4$ abundance of approximately {$7.7 \times 10^{-11}$}, and we have also found that SiH$_3$ is abundant from our hot core model (peak gas-phase abundance {$\sim 4.8 \times 10^{-11}$)}. 
SiH$_4$ exhibits an intense bending feature in the IR spectrum around 11~\micron (at 910.3 cm$^{-1}$ and 910.2 cm$^{-1}$) and has a large cross-section at 4.5~\micron that falls within the range of JWST. 
Our dark cloud model indicates that SiH$_4$, MgH$_2$, and SiCH$_2$ are significantly abundant (approximately $10^{-10}$ to $10^{-9}$) relative to H$_2$ in the ice phase, making them potentially observable species in cold dark molecular clouds for future identification with JWST. \\

\item It has been observed that HNSi and HCSi are present in detectable quantities in all the regions we studied. Additionally, the concentration of HNSi in radiation-shielded areas (such as diffuse, dark, and hot cores) is higher than that of SiN. However, while SiN has been identified, HNSi has not yet been detected. Our quantum chemical calculations, as noted in Table \ref{tab:dipole}, indicate that both species possess a negligible dipole moment, which may explain their non-detection in the gas phase. \\

\item  \cite{Gins19} proposed that NaCl could form through the reaction between ionized sodium and HCl. However, we found that this reaction is highly endothermic and would not be feasible under the conditions in a hot core. In our study, the formation of NaCl is primarily driven by the surface reaction between NaH and Cl during the initial warming phase. We also found a relatively high abundance of NaOH, approximately \(10^{-10}\), in our hot core model. Given its large dipole moment of about \(6.7\) Debye (as shown in Table \ref{tab:dipole}), NaOH could be a promising candidate for future studies. \\

\item We have identified a significant abundance of AlF in our dark cloud and hot core model through the UMIST network. However, it has not yet been observed in the regions studied here. The incorporation of ion-neutral destruction pathways significantly decreases its abundance.

\end{itemize}

\section*{Data availability}
The data underlying this article (in the appendix) are made available under a Creative Commons Attribution license on Zenodo:
doi: \href{https://doi.org/10.5281/zenodo.15727282}{10.5281/zenodo.15727282}.

\begin{acknowledgements}
    A.D. acknowledges the Max Planck Society for sponsoring a scientific visit. Part of the Gaussian computations presented in this paper were performed using the GRICAD infrastructure (\url{https://gricad.univ-grenoble-alpes.fr}), supported by Grenoble research communities. M.S. acknowledges financial support from the European Research Council (consolidated grant COLLEXISM, grant agreement ID: 811363). P.C. acknowledges the support of the Max Planck Society.
    The authors thank Arghyadeb Roy for performing part of the Gaussian computations.
    The authors are thankful to the anonymous referee for the constructive comments and suggestions, which have significantly contributed to the revision of the paper.
\end{acknowledgements}

\bibliographystyle{aa}
\bibliography{references.bib}

\begin{appendix}

\onecolumn

\section{Chemical reaction networks of metallic species with corresponding rate coefficients and enthalpies of formation for the species considered in this work.}

\begin{table}[h!]
    \centering
    \caption{The data underlying this table are made available under a Creative Commons Attribution license on Zenodo:
doi: \href{https://doi.org/10.5281/zenodo.15727282}{10.5281/zenodo.15727282}.
    \label{tab:my_label}}
\end{table}

\begin{table}[h!]
\scriptsize \centering
\caption{Enthalpies of formation for the species considered in this work. \label{tab:enthalpy}}
\begin{tabular}{cccc}
\hline
\textbf{Species}	& \textbf{$\Delta_fH^0$ (298 K) (kJ/mol)} & \textbf{Species}	& \textbf{$\Delta_fH^0$ (298 K) (kJ/mol)} \\
\hline
\multicolumn{2}{c}{\bf Na-related species} & \multicolumn{2}{c}{\bf Mg-related species} \\
\hline
  Na &	107.5 & Mg &	147.1 \\
NaH	& 124.27 & MgH	& 169.03 \\
NaO	& 83.68 & MgO &	58.16 \\
NaC	& $645.52^a$ & MgC &	$725.58^a$ \\
NaN	& $461.04^a$ & MgN &	288.7 \\
NaOH &	$-197.76$ & MgOH &	$-164.76$ \\
NaCN &	92.47  & MgCN &	$299.17^a$ \\
NaNC &	$162.67^a$ & MgNC &	$290.1^a$ \\
NaC$_2$H &	98.4 & MgC$_2$H	& $438.8^a$ \\
NaC$_4$H	& $559.9^a$ & MgC$_4$H	& $672.73^a$  \\
NaCl &	$-181.42$ & MgS	& $145.23^a$ \\
\cline{1-2}
\multicolumn{2}{c}{\bf Fe-related species} &  HMgNC &	$219.19^a$ \\
\cline{1-2}
Fe	& 415.47 & MgH$_2$	& $161.02^a$ \\
FeO	& 251.04 & MgC$_3$N	& $569.58^a$ \\
\cline{3-4}
FeS	& 370.77  &  \multicolumn{2}{c}{\bf K-related species} \\
\hline
\multicolumn{2}{c}{\bf Al-related species} & K &	89 \\
\cline{1-2}
Al &	330 & KH &	123.01  \\
AlH &	259.41 &  KCl	& $-214.68$ \\
AlO &	66.94  & KCN	& 79.5 \\
\cline{3-4}
AlF &	$-265.68$ & \multicolumn{2}{c}{\bf Si-related species} \\
\cline{3-4}
AlCl &	$-51.46$ & Si &	450 \\
AlCN &	$315.86^a$ & SiH &	376.66  \\
\cline{1-2}
\multicolumn{2}{c}{\bf Other species} &  SiO &	$-100.42$ \\
\cline{1-2}
O	& 249.18  & SiN	& 372.38  \\
H	& 218 & SiC &	719.65 \\
S &	276.98 & HCSi &	$689.47^a$ \\
Cl & 121.3 & SiH$_2$	& $271.68^a$ \\
N &	472.68 & SiH$_4$ &	34.31 \\
 O$_2$	& $2.77^a$  & SiH$_3$ &	$197.47^a$ \\
\cline{3-4}
H$_2$ &	$-1.44^a$ & \multicolumn{2}{c}{\bf Other species} \\
\cline{3-4}
H2O	& $-241.83$  & SO & 	5.01 \\
HCO	& 43.51  & OH & 	38.99 \\
HCN	& 135.14  &  CO &	$-110.53$ \\
OCN	& $124.11^a$  &  CH	& 594.13 \\
C$_2$H &	476.98  &  CN	& 435.14 \\
CH$_2$ &	386.39  & NO	& 90.29 \\
C$_4$H &	$807.74^a$  & HS	& 139.33 \\
HC$_3$N	& 354 &  CS	& 280.33 \\
HC$_5$N	& $641.29^a$ & CO$_2$ & $-393.51$ \\
SO$_2$ & $-296.81$ & H$_2$S & $-20.6$ \\
C & 716.68 & S$_2$ & 128.6 \\
SiS & 106 & & \\
\hline
\end{tabular} \\
\tablefoot{\footnotesize The values are taken from the NIST Chemistry WebBook or the SRD 69 gas-phase thermochemistry database if available. \\
{\textsuperscript{a} This work considering DFT-B3LYP/6-311+G(d,p) level of theory.}}
\end{table}

\clearpage

\section{Abundance set}

\scriptsize
{ 
    \begin{longtable}{|ccc|}
        \caption{Abundances of relevant metallic species ($> 10^{-14}$) in the PDR are presented for $A_V=5.5$ relative to total hydrogen nuclei.\label{table:pdr}} \\
        \hline
        \hline
        \textbf{Species} &  \textbf{Abundance}&\textbf{Comments}    \\
        \hline
        \endfirsthead
        \caption{\textit {continued from previous page}} \\
        \hline
       \textbf{Species} & \textbf{Abundance} & \textbf{Comments}       \\
        \hline
        \endhead
        \hline \multicolumn{3}{r}{\textit{continued on next page}} \\
        \endfoot
        \hline
        \endlastfoot
  SiH       &          2.75E-11& \\
  SiO       &          1.51E-12&{ Observed in orion bar and S 140 \citep[$10^{-11}$]{schi01b}}\\
  SiC       &          2.60E-11&\\
  SiN       &          1.02E-13&\\
  SiH+      &          6.98E-14&\\
  SiC+      &          5.12E-13&\\
  SiH2+     &          1.48E-14&\\
  HCSi      &          2.80E-12&\\
  HCSi+     &          1.63E-12&\\
  HNSi      &          3.03E-14&\\
  NaH       &          1.26E-14&\\
  MgH       &          5.87E-13&\\
  MgO       &          1.66E-14&\\
  MgH+      &          1.80E-14&\\
  FeH       &          5.89E-13&\\
  FeO       &          1.65E-14&\\
  FeH+      &          1.81E-14&\\
 \end{longtable}}

\clearpage


{     

    \begin{longtable}{|c|cc|cc|c|}
        \caption{Final abundances (after $10^6$ years with respect to H$_2$) of the metallic species for diffuse clouds {(${\rm n_H=50}$ cm$^{-3}$ and ${\rm n_H=500}$ cm$^{-3}$)} having abundance $>10^{-12}$. \label{table:diff}} \\
        \hline
        \hline
         \textbf{Species} &  \textbf{This work} & \textbf{UMIST 22}&\textbf{This work} & \textbf{UMIST 22} & \textbf{Comments} \\
                 &\multicolumn{2}{|c|}{$n_H=50$ cm$^{-3}$}&\multicolumn{2}{|c|}{$n_H=500$ cm$^{-3}$}&\\
        \hline
        \endfirsthead
        \caption{\textit {continued from previous page}} \\
        \hline
         \textbf{Species} &  \textbf{This work} & \textbf{UMIST 22}&\textbf{This work} & \textbf{UMIST 22} & \textbf{Comments}\\
                 &\multicolumn{2}{c}{$n_H=50$ cm$^{-3}$}&\multicolumn{2}{c}{$n_H=50$ cm$^{-3}$}\\       
        \hline
        \endhead
        \endfoot
        \hline
        \endlastfoot
SiH         &          2.96E-12  &          2.99E-12&          1.57E-09&          1.52E-09&\\
SiO+        &          2.59E-10  &          2.57E-10&          5.15E-10&          5.15E-10&\\
SiC+        &          4.66E-11  &          4.33E-11&          4.18E-12&          4.51E-12&\\
HCSi+       &          5.94E-12  &          5.52E-12&          3.14E-12&          3.43E-12&\\
SiN+        &          5.28E-11  &          6.08E-11&          2.19E-12&          1.05E-11&\\
SiCH2+      &          6.27E-13  &          5.90E-13&          5.39E-12&          5.97E-12&\\
HNSi+       &          8.60E-12  &          1.81E-12&          2.15E-12&          1.74E-14&\\
SiNH2+      &          6.61E-11  &          6.51E-11&          1.10E-10&          1.07E-10&\\
SiOH+       &          5.02E-10  &          4.98E-10&          1.13E-09&          1.13E-09&\\
SiF+        &          1.07E-12  &          1.06E-12&          4.45E-11&          4.45E-11&\\
SiC2+       &          3.58E-13  &          3.55E-13&          2.25E-11&          2.21E-11&\\
SiC2H+      &          6.68E-14  &          6.62E-14&          6.33E-11&          6.21E-11&\\
SiNC+       &          5.07E-11  &          5.04E-11&          4.24E-12&          1.14E-12&\\
SiC3+       &          4.90E-15  &          4.84E-15&          1.93E-11&          1.85E-11&\\
SiC4+       &          5.53E-17  &          5.46E-17&          2.24E-11&          2.19E-11&\\
SiC         &          5.22E-10  &          4.87E-10&          5.18E-08&          5.52E-08&\\
SiO         &          3.84E-08  &          3.80E-08&          1.89E-06&          1.89E-06&{ \citep[towards Sgr B2M, $3 \times 10^{-8}$]{furu03,walm02}, \citep[$(2-20) \times 10^{-11}$]{luca00}}\\
SiS         &          2.73E-13  &          3.46E-15&          1.27E-09&          7.01E-11&\\
SiH2        &          1.97E-13  &          1.97E-13&          2.43E-12&          2.38E-12&\\
SiH3        &          4.01E-13  &          3.99E-13&          3.99E-11&          3.85E-11&\\
SiH4        &          3.86E-12  &          3.83E-12&          9.01E-10&          8.84E-10&\\
HCSi        &          6.89E-11  &          6.50E-11&          2.42E-09&          2.70E-09&\\
SiN         &          9.80E-09  &          9.06E-09&          1.83E-07&          1.79E-07&{ towards Sgr B2M in absorption \citep[$\sim 10^{-7}$]{schi03}}\\
SiC2        &          1.07E-11  &          1.05E-11&          1.27E-07&          1.24E-07&\\
SiC3        &          1.46E-14  &          1.44E-14&          2.11E-08&          2.07E-08&\\
SiCH2       &          1.13E-13  &          1.13E-13&          3.03E-12&          5.60E-12&\\
HNSi        &          1.69E-08  &          1.66E-08&          1.96E-07&          1.92E-07&\\
SiCH3       &          2.20E-15  &          2.17E-15&          1.58E-12&          3.18E-12&\\
SiC2H2      &          1.59E-17  &          1.56E-17&          3.75E-12&          3.79E-12&\\
SiO2        &          2.50E-17  &          2.50E-17&          3.76E-12&          3.60E-12&\\
NaH         &          7.22E-14  &          5.15E-10&          4.99E-13&          5.11E-09&{ along $\zeta$ Ophiuchi \citep[$<1.3 \times 10^{10}$ cm$^{-2}$]{czar87}}\\
NaCN        &          4.03E-20  &          1.82E-18&          7.59E-13&          4.53E-09&\\
NaC9NH+     &          7.07E-24  &          6.85E-24&          1.01E-12&          1.01E-12&\\
MgH         &          5.66E-12  &          4.83E-11&          2.12E-10&          9.08E-10&{along $\zeta$ Ophiuchi \citep[$<1.9 \times 10^9$ cm$^{-2}$]{czar87}}\\
MgH2        &          2.31E-11  &          1.01E-08&          1.23E-09&          1.35E-07&\\
MgO         &          2.45E-13  &          6.10E-13&          1.56E-11&          6.55E-12&\\
MgCN        &          3.44E-18  &          1.67E-16&          9.60E-19&          2.96E-12&\\
MgNC        &          6.90E-18  &          3.42E-16&          4.78E-12&          2.85E-08&\\
MgS         &          3.22E-16  &                 -&          3.05E-12&                -&\\
HMgNC       &          1.17E-22  &          3.18E-20&          4.78E-14&          2.84E-10&\\
MgC5N       &          7.63E-21  &          7.48E-21&          3.18E-10&          3.13E-10&\\
MgC7N       &          6.85E-21  &          6.65E-21&          9.81E-09&          9.54E-09&\\
MgC2        &          7.66E-19  &          7.52E-19&          1.58E-06&          1.55E-06&\\
MgC2H       &          3.14E-18  &          3.09E-18&          1.57E-06&          1.54E-06&\\
MgC6H       &          4.74E-19  &          4.63E-19&          8.49E-07&          8.10E-07&\\
MgC8H       &          8.61E-18  &          8.37E-18&          1.48E-05&          1.45E-05&\\
MgC7NH+     &          2.94E-22  &          2.86E-22&          3.75E-12&          3.65E-12&\\
MgC6H2+     &          3.53E-22  &          3.45E-22&          1.67E-11&          1.59E-11&\\
MgC8H2+     &          7.23E-21  &          7.04E-21&          2.90E-10&          2.85E-10&\\
AlO         &          3.03E-14  &          3.52E-13&          2.75E-10&          4.45E-10&\\
AlF         &          4.68E-15  &          1.52E-14&          1.14E-11&          1.80E-11&\\
FeH         &          1.53E-12  &          2.52E-09&          9.17E-10&          6.07E-08&\\
FeO         &          2.01E-14  &          2.13E-22&          2.66E-10&          7.54E-20& { 1 transitions towards UC-HII-Sgr B2M in absorption \citep[$3 \times 10^{-11}$]{furu03,walm02}}\\    
 \end{longtable}}

{ 
    \begin{longtable}{|cccc|}
        \caption{Final abundances (after \(10^6\) years relative to H$_2$) of metallic species in dark clouds, exhibiting either abundances greater than \(10^{-12}\) or targeted upper limits available. 
        \label{table:dense}} \\
       \hline
        \hline
        \textbf{Species} &  \textbf{This work} & \textbf{UMIST 22}& \textbf{Comments}   \\
        \hline
        \endfirsthead
        \caption{\textit {continued from previous page}} \\
        \hline
       \textbf{Species} & \textbf{This work} & \textbf{UMIST 22}    & \textbf{Comments}  \\
        \hline
        \endhead
        \hline \multicolumn{3}{r}{\textit{continued on next page}} \\
        \endfoot
        \hline
        \endlastfoot
SiNH2+      &          1.38E-12  &          1.06E-12&\\
SiOH+       &          2.91E-11  &          3.10E-11&\\
SiO         &          3.65E-09  &          3.92E-09& { \citep[TMC-1, $<2.4 \times 10^{-12}$]{ziur89}}\\

SiS         &          2.64E-12  &          3.26E-12&{ \cite[TMC-1. $<5.9 \times 10^{-11}$]{ziur91}}\\
SiH3        &          5.44E-11  &          5.78E-11&\\
SiH4        &          4.83E-10  &          5.15E-10&\\
HCSi        &          6.33E-12  &          6.66E-12&\\
SiN         &          5.58E-11  &          4.34E-11&\\

SiCH2       &          1.09E-11  &          1.17E-11&\\
HNSi        &          8.49E-10  &          6.95E-10&\\
SiO2        &          9.04E-12  &          9.69E-12&\\
NaH         &          4.99E-13  &          9.16E-12&\\
MgH         &          2.84E-12  &          4.34E-12&\\
MgH2        &          9.39E-12  &          2.45E-11&\\
MgC8H       &          1.01E-12  &          1.13E-12&\\
AlO         &          2.07E-12  &          4.10E-11&\\
MgNC        &          5.09E-15  &          8.68E-13&  { TMC-1 \citep[$<1.2 \times 10^{-11}$]{turn05} }\\
AlF         &          1.40E-13  &          2.67E-10&\\
FeH         &          1.39E-12  &          1.01E-11&\\
\end{longtable}}

{ 
    \begin{longtable}{|cccc|}
        \caption{Peak abundances (in between the warm-up and post-warm-up stage with respect to H$_2$) of the metallic species for a hot core having abundance $>10^{-12}$.  \label{table:HC}} \\
       \hline
        \hline
        \textbf{Species} &  \textbf{This work} & \textbf{UMIST 22} & \textbf{Comments}   \\
        \hline
        \endfirsthead
        \caption{\textit {continued from previous page}} \\
        \hline
       \textbf{Species} & \textbf{This work} & \textbf{UMIST 22}  &  \textbf{Comments}  \\
        \hline
        \endhead
        \hline \multicolumn{3}{r}{\textit{continued on next page}} \\
        \endfoot
        \hline
        \endlastfoot
SiH         &          7.19E-12  &          8.14E-12& { Tentative in Orion KL \citep[$3 \times 10^{-9}$]{schi01}} \\
SiO         &          3.40E-10  &          3.44E-10&
{ Tentative in Orion KL \citep[$1.2 \times 10^{-9}$]{wils71,ziur91}, \citep[Orion-KL, $\sim 2.0 \times 10^{-7}$]{ziur89}}\\
SiS         &          2.71E-11  &          3.54E-11&\\
SiH2        &          6.59E-12  &          7.31E-12&\\
SiH3        &          5.09E-11  &          5.84E-11&\\
SiH4        &          7.38E-11  &          7.96E-11&\\
HCSi        &          5.83E-12  &          6.72E-12&\\
SiN         &          4.19E-11  &          3.19E-11&\\
SiC2        &          1.12E-11  &          7.77E-12&\\
SiCH2       &          5.06E-12  &          6.39E-12&\\
HNSi        &          1.12E-10  &          1.05E-10&\\
H2SiO       &          1.46E-11  &          1.63E-11&\\
SiC2H       &          2.13E-12  &          2.03E-12&\\
SiC2H2      &          3.63E-12  &          4.01E-12&\\
SiO2        &          3.71E-12  &          3.97E-12&\\
NaH         &          1.07E-09  &          1.78E-09&\\
NaOH        &          1.71E-10  &          -&\\
NaO         &          5.41E-11  &          -&{ Tentative in Orion KL \citep[$<1.6 \times 10^{-9}$]{turn91}}\\
NaCN        &          1.04E-14  &          1.27E-12&\\
NaCl        &          2.56E-11  &          2.43E-11&{\cite{Gins19,wrig20}}\\
MgH         &          4.51E-11  &          1.88E-12&\\
MgH2        &          4.41E-10  &          4.87E-10&\\
MgCN        &          5.43E-17  &          4.66E-12&\\
MgNC        &          2.98E-15  &          9.58E-12&\\
MgC7N       &          1.49E-12  &          3.96E-14&\\
MgC2        &          4.30E-11  &          1.43E-11&\\
MgC2H       &          1.91E-12  &          4.82E-14&\\
MgC8H       &          7.91E-12  &          2.01E-13&\\
AlO         &          3.42E-10  &          1.65E-11&\\
AlH         &          6.30E-12  &          4.78E-14&\\
AlF         &          4.06E-11  &          3.81E-10&\\
AlOH        &          2.70E-11  &          2.95E-12&\\
Al2O        &          8.14E-12  &          4.80E-13&\\
FeH         &          9.49E-10  &          9.67E-10&\\
FeO         &          1.19E-12  &          8.42E-16&{ \citep[Orion KL, $<10^{-11}$]{mere82}}\\
KH          &          8.60E-11  &          -&\\
\end{longtable}}

\end{appendix}

\end{document}